\documentclass[10pt]{article}
\usepackage[T1]{fontenc} 
\usepackage[cp1250]{inputenc}
\usepackage{graphicx} 
\usepackage[english]{babel} 
\usepackage{psfrag}
\usepackage[normalem]{ulem}
\usepackage{authblk}
\usepackage{amssymb, amsmath, amsfonts}
\usepackage{bm}
\usepackage{multirow}
\usepackage{graphicx}
\usepackage{mathtools}
\usepackage{subfig}
\usepackage{csquotes}
\include{subfig}
\usepackage[normalem]{ulem} 
\usepackage[usenames,dvipsnames]{color}

{\begin{list}{}%
         {\setlength{\leftmargin}{#1}}%
         \item[]%
}
{\end{list}}

\newcommand{\bq}{\begin{equation}}
\newcommand{\eq}{\end{equation}}
\newcommand{\ba}{\begin{eqnarray}}
\newcommand{\ea}{\end{eqnarray}}

\author[1]{Bartlomiej Waclaw}
\author[2]{Justyna Cholewa-Waclaw}
\author[3,4]{Philip Greulich}
\affil[1]{School of Physics and Astronomy,  University of Edinburgh, Edinburgh, UK}
\affil[2]{School of Biological Sciences, University of Edinburgh, Edinburgh, UK}
\affil[3]{School of Mathematics, University of Southampton, Southampton, UK}    
\affil[4]{Institute for Life Sciences, University of Southampton, Southampton, UK} 
\begin{document}
\title{Totally asymmetric exclusion process with site-wise dynamic disorder}

\date{\today}

\maketitle 
\noindent
\begin{abstract}
We propose an extension of the totally asymmetric simple exclusion process (TASEP) in which particles hopping along a lattice can be blocked by obstacles that dynamically attach/detach from lattice sites. The model can be thought as TASEP with site-wise dynamic disorder. We consider two versions of defect dynamics: (i) defects can bind to any site, irrespective of particle occupation, (ii) defects only bind to sites which are not occupied by particles (particle-obstacle exclusion). In case (i) there is a symmetric, parabolic-like relationship between the current and particle density, as in the standard TASEP. Case (ii) leads to a skewed relationship for slow defect dynamics. We also show that the presence of defects induces particle clustering, despite the translation invariance of the system. For open boundaries the same three phases as for the standard TASEP are observed, albeit the position of phase boundaries is affected by the presence of obstacles. We develop a simple mean-field theory that captures the model's quantitative behaviour for periodic and open boundary conditions and yields good estimates for the current-density relationship, mean cluster sizes and phase boundaries. Lastly, we discuss an application of the model to the biological process of gene transcription.
\end{abstract}

\section{Introduction}
\label{intro_sec}

Consider an every-day scenario of cars travelling down a road. 
If the density of cars is low, cars travel smoothly and there is no congestion. However, a high density of cars or the presence of obstacles (e.g. traffic lights) can induce queuing of vehicles which leads to a congested state in which traffic slows down or even comes to a halt.
Similar scenarios also occur in the microscopic world of molecular biology. There, ``vehicles'' can be molecular motors proceeding along intracellular filaments or DNA/mRNA strands, or ions migrating through ion channels.

The basic features of those situations are captured by the totally asymmetric simple exclusion process (TASEP) \cite{spitzer_interaction_1970}, the paradigmatic model for stochastic transport in which particles may hinder each other's movement. In its simplest incarnation, TASEP describes a system of particles hopping unidirectionally between the sites of a one-dimensional open lattice. Only one particle can occupy a given site at a time. This excluded-volume effect leads to particles colliding with each other, causing congestion when the particle density is sufficiently high.

TASEP was originally proposed to model biopolymerization such as the synthesis of RNA on DNA templates \cite{macdonald_kinetics_1968} but since then TASEP and related models have been applied to a variety of phenomena: protein production \cite{chou1,multi-tasep}, traffic flow \cite{traffic1,andreas_traffic2}, the movement of molecular motors \cite{pff1,lipowsky_2001,lipowski_network}, transport through ion channels \cite{kolomeisky_channel-facilitated_2007}, and pedestrian traffic \cite{traffic1}. From the theory standpoint, TASEP has been extensively studied as an archetype model of jamming \cite{spitzer_interaction_1970,Derrida1998,chou_non-equilibrium_2011,andreas_traffic2}, helped by the property that it is exactly solvable and that mean-field approach gives the same result as the exact solution \cite{derrida_exact_1993}. 
A celebrated property of the TASEP with open boundary conditions, when particles enter the lattice from a reservoir at one end and exit at the other, is the existence of phase transitions, even though it is a one-dimensional system \cite{krug1}. These phase transitions, between a low density, high density and a maximum current phase, are reminiscent of particle queueing and congestion observed in traffic-like systems as those mentioned above.

In real-world transport, movement is often hindered by obstacles. On a road this may be crossings or traffic lights, while molecular traffic is often impeded by bound proteins or some transient modifications of the ``lane'' on which traffic occurs. For example, when mRNA is synthesized by RNA polymerase from a DNA template in the process of transcription \cite{alberts_molecular_2002}, the polymerase encounters ``roadblocks'' that slow down its progress. These roadblocks can be any DNA-bound structural and regulatory proteins that must be removed for the polymerase to proceed, for example histones that form the core of nucleosomes \cite{li_role_2007}.

Such obstacles -- which we shall call \emph{defects} -- have been extensively studied in the context of TASEP, when specific sites or bonds have a hopping rate that differs from others. In this context one can consider single defects \cite{lebowitz_1def_1,1def_kolomeisky}, or \emph{quenched site-wise disorder}, the random distribution of spatially varying hopping rates \cite{barma1,barma2,barma_driven_2006,PASEP_dis_ludger,asep_def,asep_dis}. Although no exact solution of the TASEP with defects exists\footnote{In contrast, for the situation of particle-wise disorder, in which the hopping rate varies between particles, exact solutions are possible \cite{DTASEP_krug}.}, these studies have significantly improved our understanding of transport with obstacles, and have exemplified the hallmark of such systems: a phase separation (queuing of particles) even in periodic systems, and a reduction of the carrying capacity in open systems. 

Recently, models with dynamic defects have been studied.
For example, transcription "roadblocking" has been considered in computational biology literature. Computer simulations of a single dynamic roadblock were able to explain the behaviour of {\it E. coli lac} repressor (LacI) \cite{hao_road_2014}. A more complex model involving cooperation between polymerases in removing a roadblock has been applied to explain why transcription is not significantly compromised in the presence of DNA-bound proteins  \cite{epshtein_transcription_2003}. Non-biological applications include periodically switching traffic lights \cite{ludger_many-dyn-defs}, obstacles that stochastically move and perform long range hops \cite{klumpp_dyn-def_2014}, or obstacles that bind and unbind stochastically to specific sites \cite{turci_transport_2013,klumpp_dyn-def_2016}. Inherent to these models is that similarly to the static defect case, the translational invariance is broken and thus congestion occurs at defect sites. In fact, for fast defect dynamics, these systems behave very similar to static defect systems.

The situation of random dynamic defects where defects can bind to any site has been less studied. Results exist for the partially asymmetric exclusion process (with hopping in opposite direction allowed) without average bias, in which the locally preferred transport direction varies dynamically \cite{barma_driven_2006} and in the totally asymmetric case for the ``bus-route model'' \cite{bus-route}, where defects appear randomly, but are removed by particles. In the former work, however, the model is globally symmetric (and not totally asymmetric locally), while in the latter the particle-defect interaction introduces an additional feedback that makes it difficult to identify the plain effect of defects.


Here we propose a simple process, a TASEP with \emph{dynamic disorder} (ddTASEP), in which defects appear and disappear randomly and uniformly across the lattice, and when present, slow-down or stop particles from moving down the chain of sites. The defects are thought of as obstacles that bind and unbind from an infinite reservoir. In contrast to previous instances of dynamic defects, this system is fully asymmetric, retains translational invariance and, in its basic version, defect dynamics is independent of particle occupation. We will also consider a version in which obstacles and particles are mutually exclusive.

To explore the dynamics of the model, we will first simulate the model on a computer to obtain the \emph{current-density relation (CDR)} -- the relationship between the current $J$ of particles and particle density $\rho$. We will study what effect the dynamic disorder has on the CDR and how it depends on the density of defects and the timescale of defect turnover. For that purpose, we will develop a mean field approach which captures the main features of the CDR and provides a reasonable estimate for the current. We shall see that despite the preserved translational invariance of the system, the distribution of particles exhibits a high degree of inhomogeneity, and we will present a theory for the formation of particle clusters which is able to provide a good estimate for the mean cluster size. These results will be used to predict the effect of dynamic disorder on the phase diagram of the TASEP for open boundary conditions. Finally, we will show that our model can be used to explain some aspects of the global regulation of gene expression.

\section{Model}
\label{model_sec}
\subsection{Definition of model dynamics}
\label{model_def_sec}

We consider a totally asymmetric simple exclusion process \cite{spitzer_interaction_1970} with \emph{{\bf d}ynamic {\bf d}isorder} (ddTASEP), in which defects that slow down particles can appear and disappear on any site. Particles and defects reside on sites $i=1,...,L$ of a one-dimensional lattice. A particle hops from site $i$ to $i+1$ with rate $p$ if the arrival site $i+1$ is empty. If the arrival site contains a defect, the particle hops with rate $p_d<p$. If $p_d=0$, the defect can be thought as representing a physical obstacle blocking the particle. Defects appear and disappear stochastically: A site without a defect acquires a defect with rate $k_+$, whereas a defect site switches to a non-defect site with rate $k_-$.
Motivated by biological scenarios, we refer to this transition as defect binding/unbinding, respectively.

We consider two variants of the model. In the \emph{unconstrained} version 
defects can bind to a site without any restriction. If $\sigma_i=0,1$ denotes the absence/presence of a particle at site $i$, $\nu_i=0,1$ is the absence/presence of a defect, and $p_i$ is the hopping rate $i\to i+1$, we can formally write the model dynamics for bulk sites, $1 < i < L$, as
\begin{align}
\sigma_{i}=1,\sigma_{i+1}=0 &\xrightarrow{p_{i}} \sigma_{i}=0,\sigma_{i+1}=1, \label{eq:model1} \\
\nu_{i} =1 &\xrightarrow{k_-} \nu_{i} =0, \label{eq:model2} \\
\nu_{i} =0 &\xrightarrow{k_+} \nu_{i} =1, \label{eq:model3}
\end{align}
where
\bq
p_{i} = \left \lbrace \begin{array}{ll}
p & \mbox{ if } \nu_{i+1} = 0, \\ \nonumber
p_{d} & \mbox{ if } \nu_{i+1} = 1,
\end{array}\right.  
\eq
with $p_d < p$. Note that, in general, $p_{i} = p_{d} \nu_{i+1} + p (1 - \nu_{i+1})$, thus we can equivalently write the defect dynamics from Eqs. (\ref{eq:model2},\ref{eq:model3}) as $ p_{i} = p_{d} \xrightleftharpoons[k_+]{k_-} p$.

We shall further consider two types of boundary conditions (BC): \emph{periodic BC} (see Section \ref{periodic_sec}), for which particles at site $L$ re-enter at site $1$ when hopping, 
\begin{align}
\sigma_{L}=1,\sigma_{1}=0 &\xrightarrow{p_{L}} \sigma_{L}=0,\sigma_{1}=1, \label{eq:model_per}
\end{align}
and \emph{open BC} (see Section \ref{open_bc_sec}), for which at sites $1$ and $L$ particles enter/exit from a reservoir, respectively, with rates $\alpha$ and $\beta$,
\begin{align}
\sigma_{1}=0 &\xrightarrow{\alpha} \sigma_{1}=1 , \\ 
\sigma_{L}=1 &\xrightarrow{\beta} \sigma_{L}=0 ,
\label{eq:model_open}
\end{align}
in addition to normal particle hopping, Eq. (\ref{eq:model1}), on site $1$. The defect binding dynamics are the  same in the bulk and on boundaries. 

In the \emph{constrained} version, a defect can only bind if the respective site is not occupied by a particle. Equation (\ref{eq:model3}) is then replaced by
\begin{align}
\label{model_varB}
\nu_{i} = 0 \xrightarrow{k_+(1-\sigma_{i})} \nu_{i} = 1, 
\end{align}
while Eqs. (\ref{eq:model1}-\ref{eq:model2}) remain unchanged. 

If not specified otherwise in the text, we consider the former, unconstrained variant, Eqs. (\ref{eq:model1}-\ref{eq:model3}). 

\begin{figure*}
	\includegraphics[angle=270,width=\columnwidth]{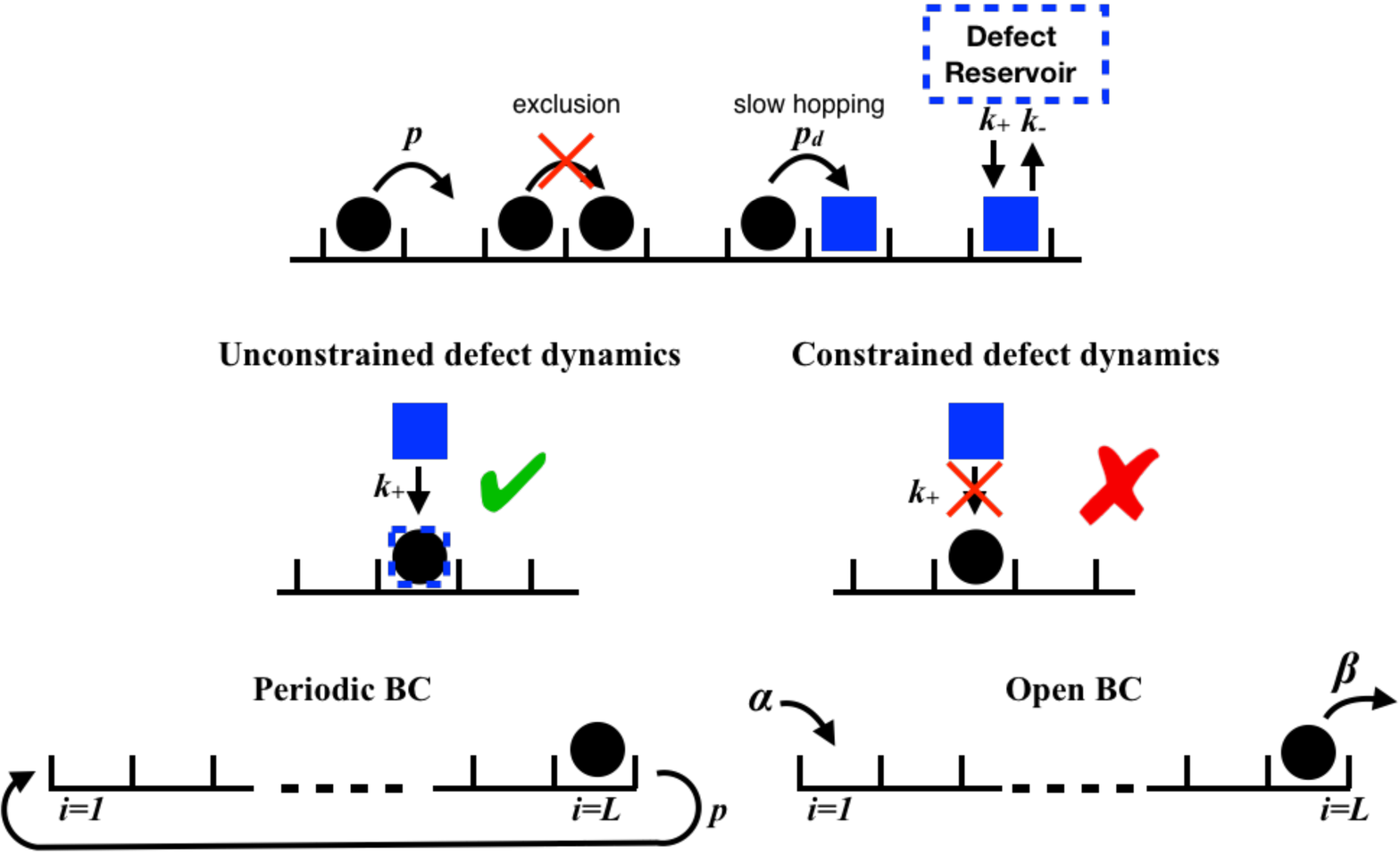}
	\caption{\label{model_fig} Illustration of the model (see Section \ref{model_def_sec} for description).}
\end{figure*}

The model dynamics are illustrated in Fig. \ref{model_fig}. Equations (\ref{eq:model1}-\ref{eq:model3}) define the model via a set of chemical-like reactions. This description is convenient if one wants to study the behaviour of the model numerically. Here we use a Monte Carlo algorithm with random sequential update. A single step of the algorithm consists of choosing a random site, selecting an event (a particle attempts to hop, a defect binds/unbinds) with probability proportional to the rates (\ref{eq:model1}-\ref{eq:model3}), and increasing the time variable by $1/\max(p+k_-,p+k_+)$. Although not exact, this algorithm is faster than the exact kinetic Monte Carlo method (Gillespie algorithm \cite{gillespie}) and the results quickly converge to the exact results for $L\gg 1$.

\subsection{Observables}
	
A crucial observable in the TASEP and related models is the mean particle current $J$ in the steady state. Let us first define the average current $J_i$ at site $i$ as the total rate at which particles hop across the bond $(i,i+1)$. In the ddTASEP, a hop occurs with probability $p_{i}$ whenever site $i$ is occupied and site $i+1$ is not occupied, therefore
\bq
J_{i} = \langle p_{i} \sigma_{i}(1-\sigma_{i+1}) \rangle \,\,\,.
\eq
We note that $J_i$ is the inverse of the mean waiting time $\bar \tau$ per particle, so that the steady state particle current can be also defined as
\bq
\label{J_i_def_eq}
J_{i} = \bar \tau^{-1} \langle \sigma_i \rangle .
\eq
At steady state, $J_1=J_2=\dots=J_N =: J$ due to the local conservation of particles. In the case of periodic boundary conditions, we are  in particular interested in the relationship between current $J$ and particle density $\rho := \langle \sigma_i \rangle$, i.e., the function $J(\rho)$, also called the \emph{current-density relation (CDR)}. For the standard TASEP,
\bq
\label{eq:jvsrho_standard}
J(\rho) = p\rho(1-\rho),
\eq
which is an inverted parabola, with maximum $J_{\rm max}:= \max_\rho[J(\rho)]=1/4$ at $\rho_{\rm max}=1/2$ \cite{derrida_exact_1993}. Another quantity of interest is the correlation between particle occupancies at neighbouring sites, 
\bq
	C(\sigma_{i},\sigma_{i+1}) := \langle \sigma_{i} \sigma_{i+1} \rangle - \langle \sigma_{i} \rangle \langle \sigma_{i+1} \rangle .\label{eq:corr}
\eq
which is an estimate for the deviation of typical mean field approaches from exact results.

\section{Periodic Boundary Conditions}
\label{periodic_sec}

We first consider the ddTASEP on a lattice of $L$ sites with periodic boundary conditions, according to Eq. (\ref{eq:model_per}). In this case the total number of particles $N$ is conserved. We are interested in the limit of $L,N\to\infty$ and fixed density of particles $\rho = N/L$. 

\subsection{Unconstrained defect dynamics, full-blocking defects ($p_{d}=0$)}
\label{unconstrained_sec}

\begin{figure}[h]
	\subfloat[]{\includegraphics[width=0.48\columnwidth]{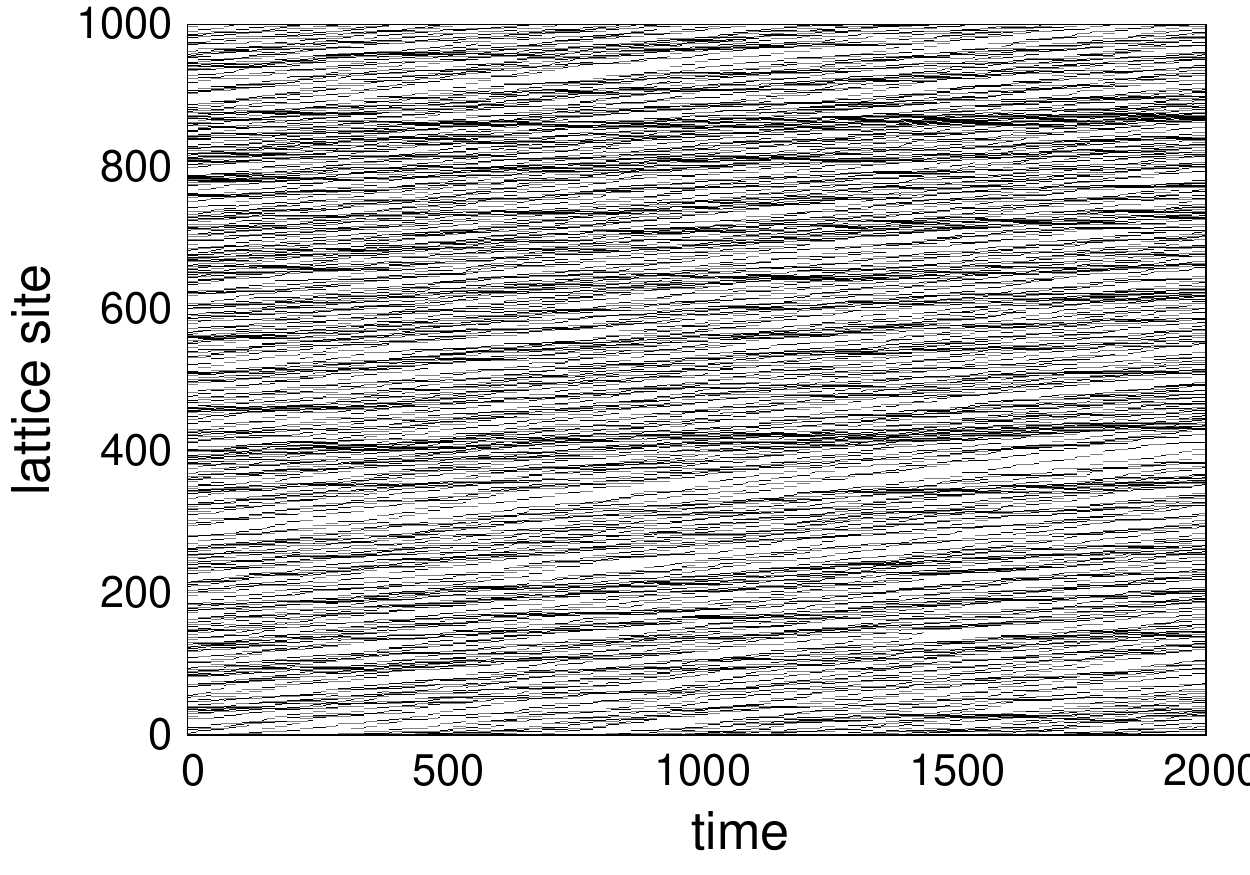}}
	\subfloat[]{\includegraphics[width=0.48\columnwidth]{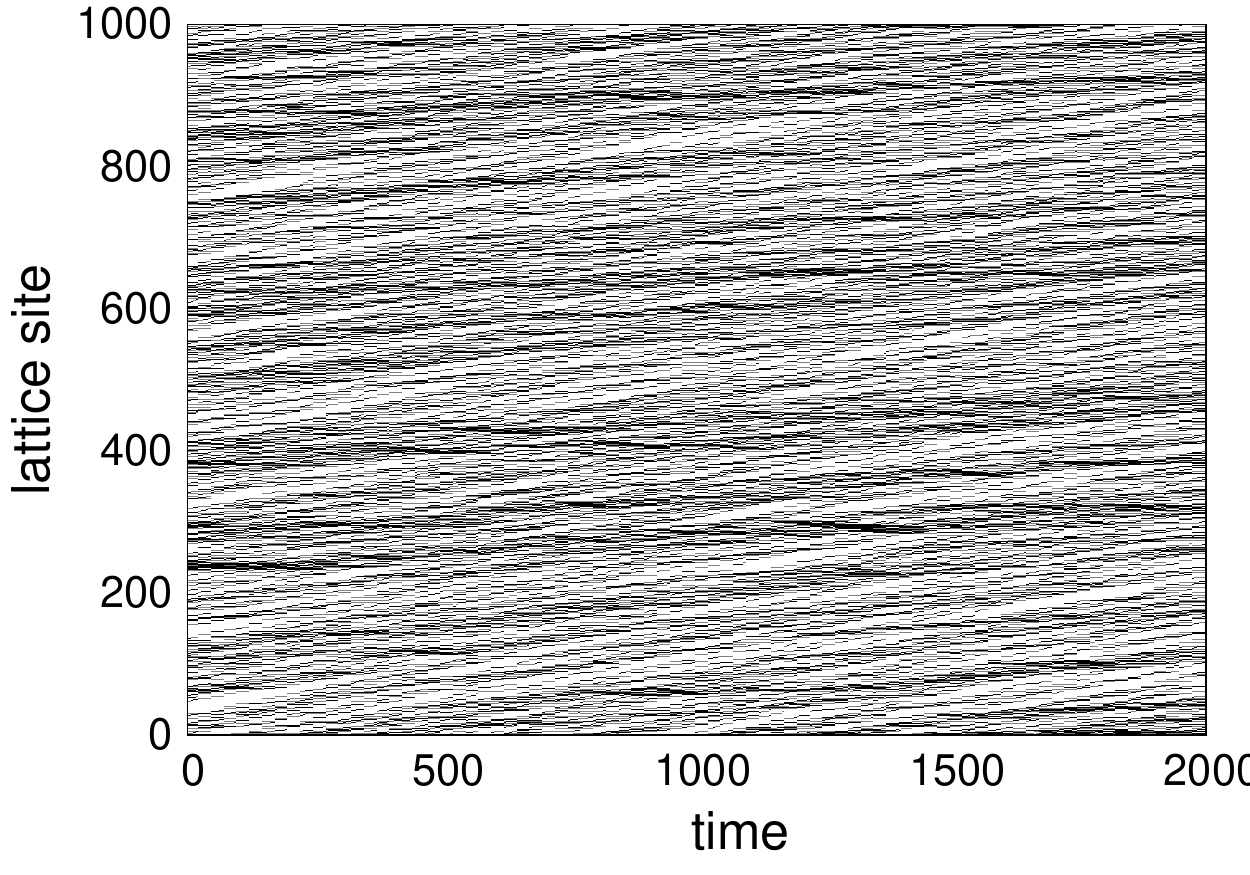}}\\
	\subfloat[]{\includegraphics[width=0.48\columnwidth]{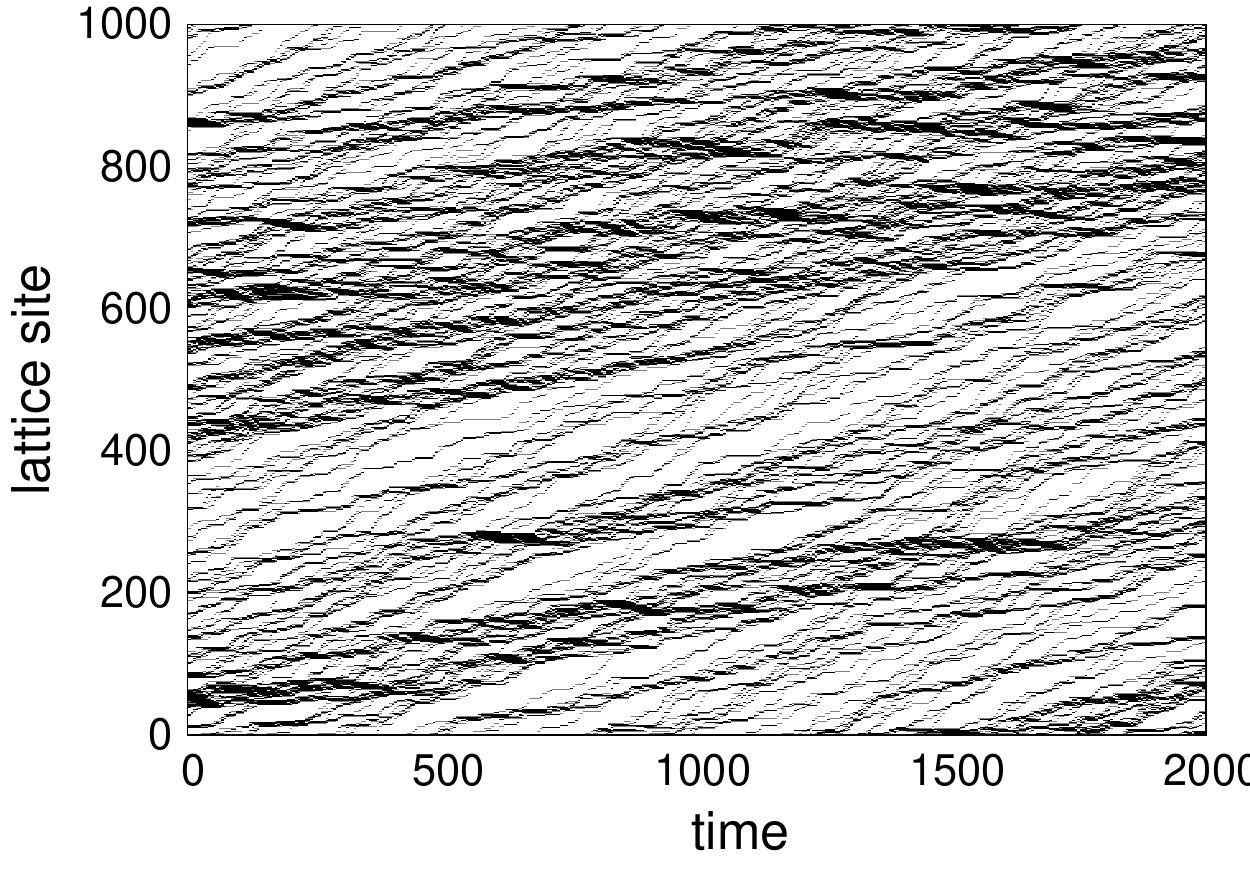}}
	\subfloat[]{\includegraphics[width=0.48\columnwidth]{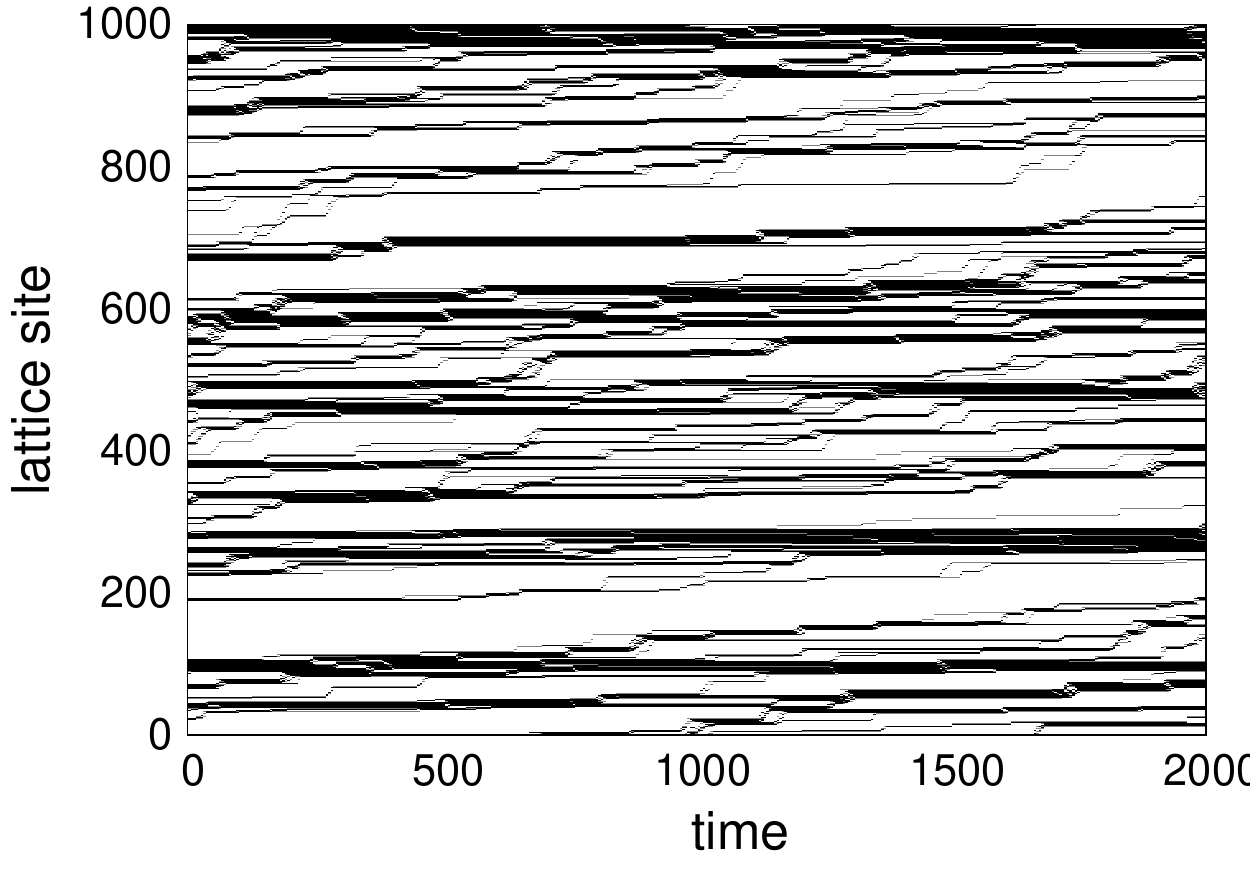}}
	\caption{\label{space-time_plot} Space-time plot for $\rho=0.3$; each pixel denotes a particle where the $y-$axis denotes the lattice site and the $x-$axis time, in units of $p^{-1}$.  (a) $k_- = 5p,\, k_+ = 5p$. (b) $k_- = 5p,\, k_+ = p$. (c) $k_- = 0.1p,\, k_+ = 0.02p$. (d) $k_- = 0.01p,\, k_+ = 0.002p$.}
\end{figure}

\begin{figure}[h]
\subfloat[]{\includegraphics[width=0.48\columnwidth]{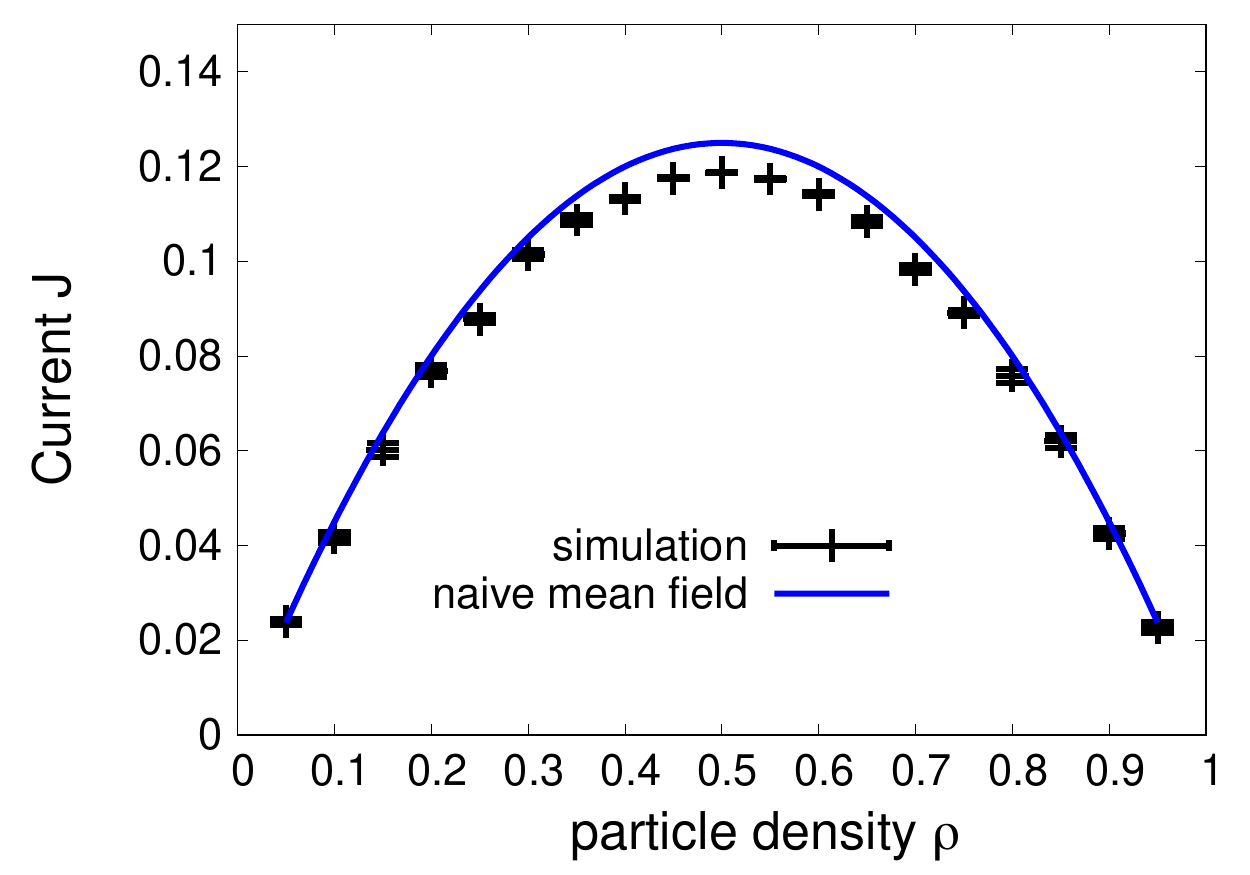}}
\subfloat[]{\includegraphics[width=0.48\columnwidth]{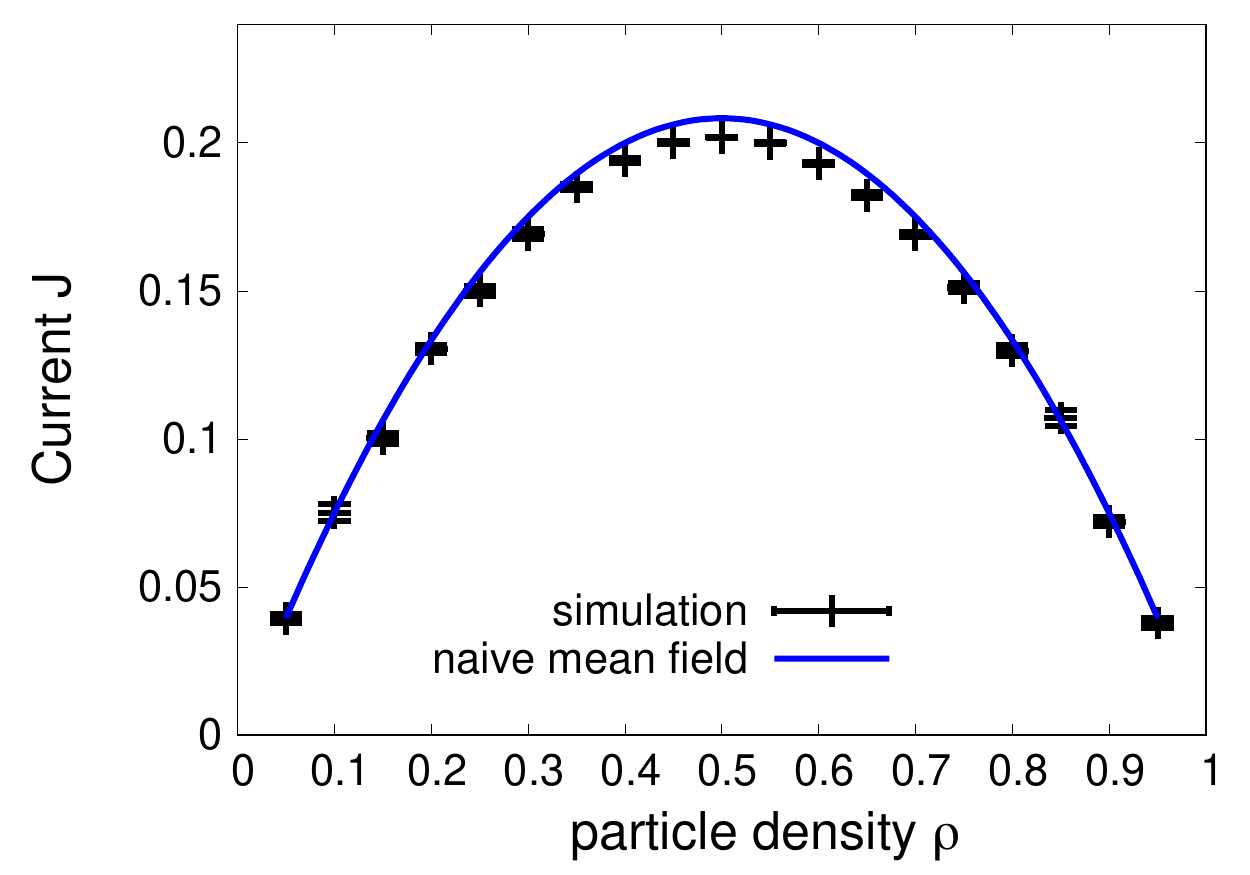}}\\
\subfloat[]{\includegraphics[width=0.48\columnwidth]{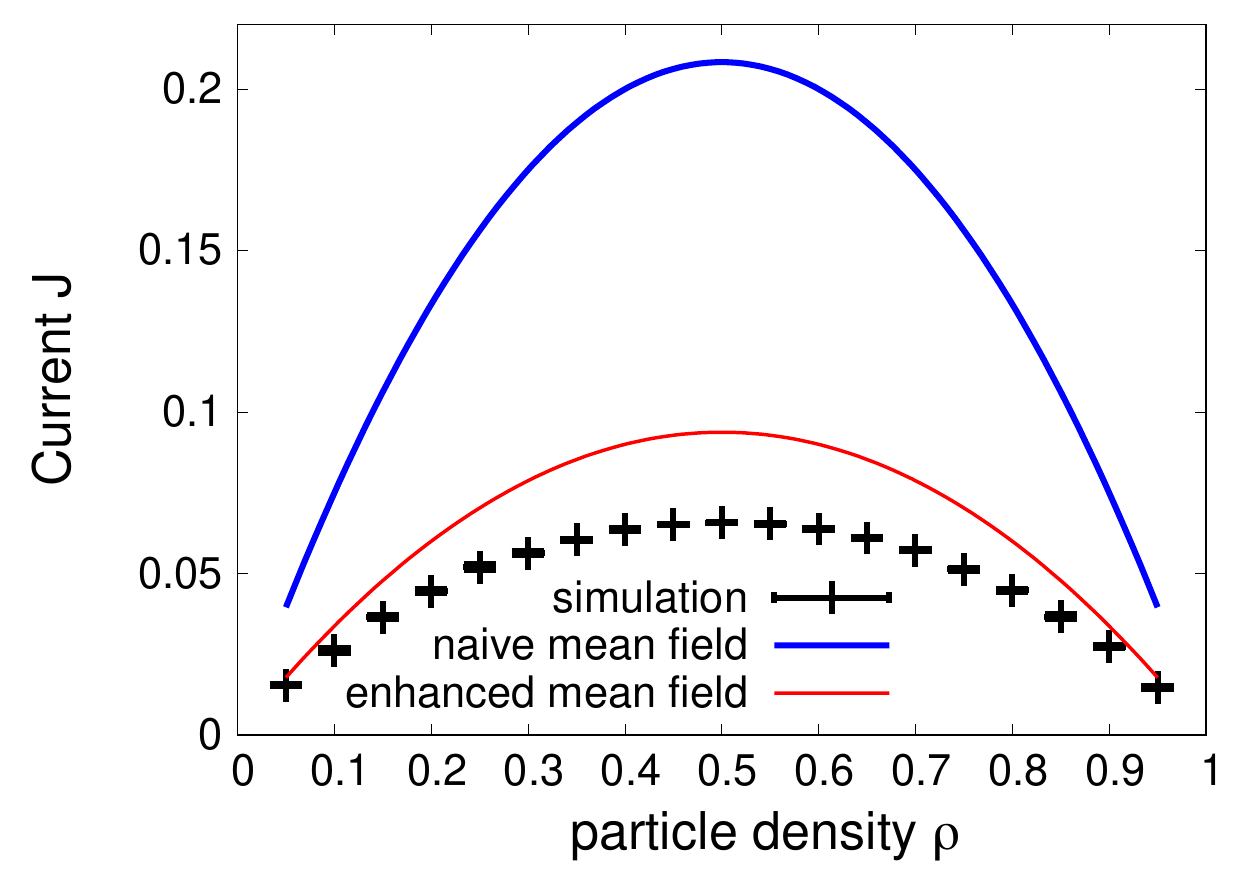}}
\subfloat[]{\includegraphics[width=0.48\columnwidth]{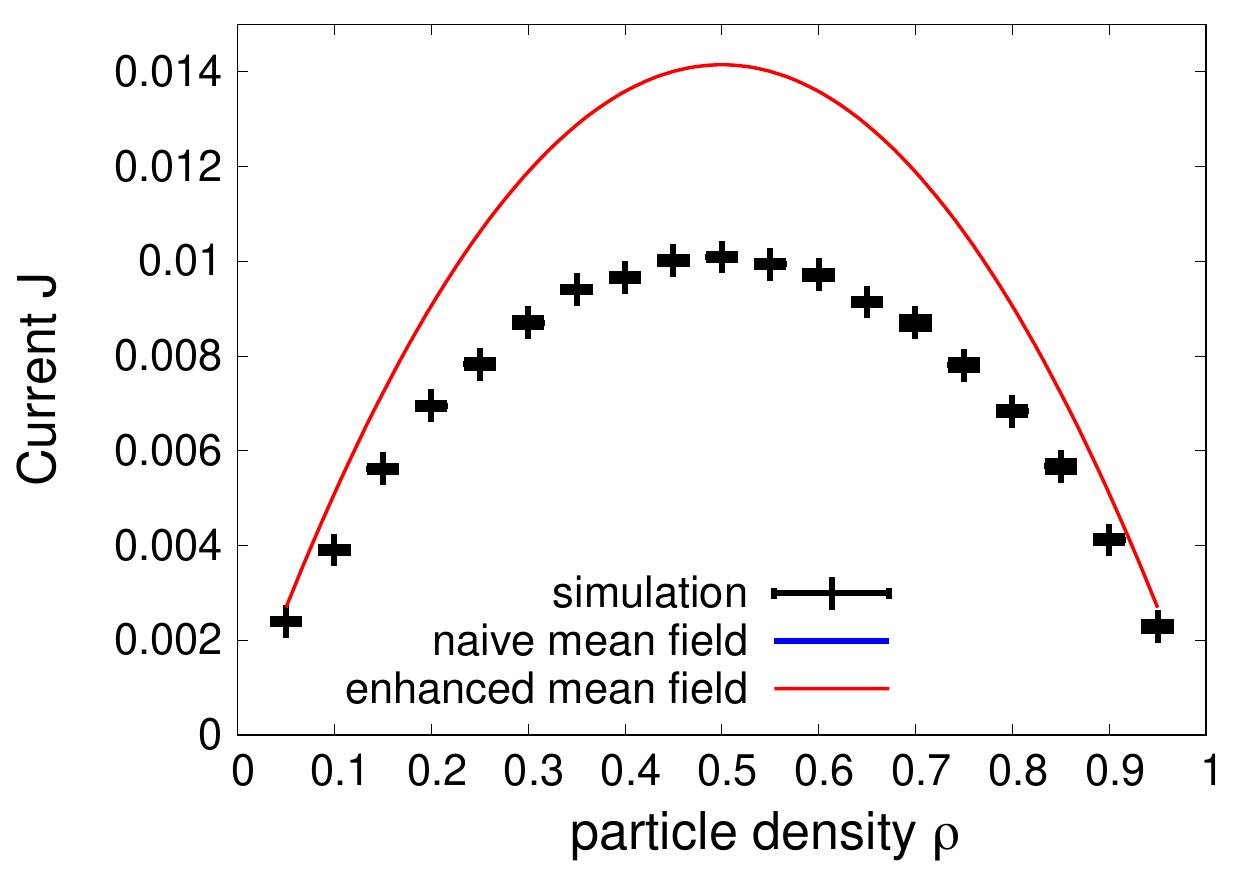}} \\
\caption{\label{J_rho_fig} Particle current $J$, in units of $p$, as a function of particle density $\rho = \langle \sigma_{i} \rangle$, for $T=100000/p$, $L=1000$ and different rates of defect binding/unbinding rates $k_{+,-}$. The blue line is the naive mean field approximation (Eq. (\ref{naive_mf_eq})), the red line is the enhanced mean field approximation (Eq. (\ref{enhanced_mf_eq})). (a) $k_- = 5p, k_+ = 5p$. (b) $k_- = 5p, k_+ = p$. (c) $k_- = 0.1p, k_+ = 0.02p$. (d) $k_- = 0.01p, k_+ = 0.002p$. Error bars (if not visible, they are smaller than the symbol size) are standard error of mean for 10 replicates.}
\end{figure}

We assume that defects block particle hopping entirely, so that the hopping rate in the presence of the defect is $p_{d}=0$, and that binding of a defect is independent of the particle occupation of a site (Eqs. (\ref{eq:model1} - \ref{eq:model3})). Figure \ref{space-time_plot} shows the space-time plots obtained by computer simulations for different values of the defect binding/unbinding rates and density $\rho=0.3$ which in the standard TASEP would lead to smooth (non-congested) flow. Indeed, particles are uniformly distributed over the lattice for high binding/unbinding rates. However, as the rates decrease and defects stay longer on the lattice, particles begin to cluster. This is reflected in the CDR (Fig. \ref{J_rho_fig})\footnote{Note that in all figures where error bars are displayed, and where it is not further specified, the error bars denote the standard error of mean of 10 replicate simulation runs.}. A characteristic parabolic shape resembling the CDR of the standard TASEP (Eq. (\ref{eq:jvsrho_standard})), can be observed. However, the maximum current is reduced compared to the TASEP maximum current, $J_{\rm max}|_{p_d = 1} = 1/4$, and decreases with decreasing $k_+,k_-$ (different panels of Fig. \ref{J_rho_fig}).

\begin{figure}[h]
\subfloat[]{\includegraphics[width=0.48\columnwidth]{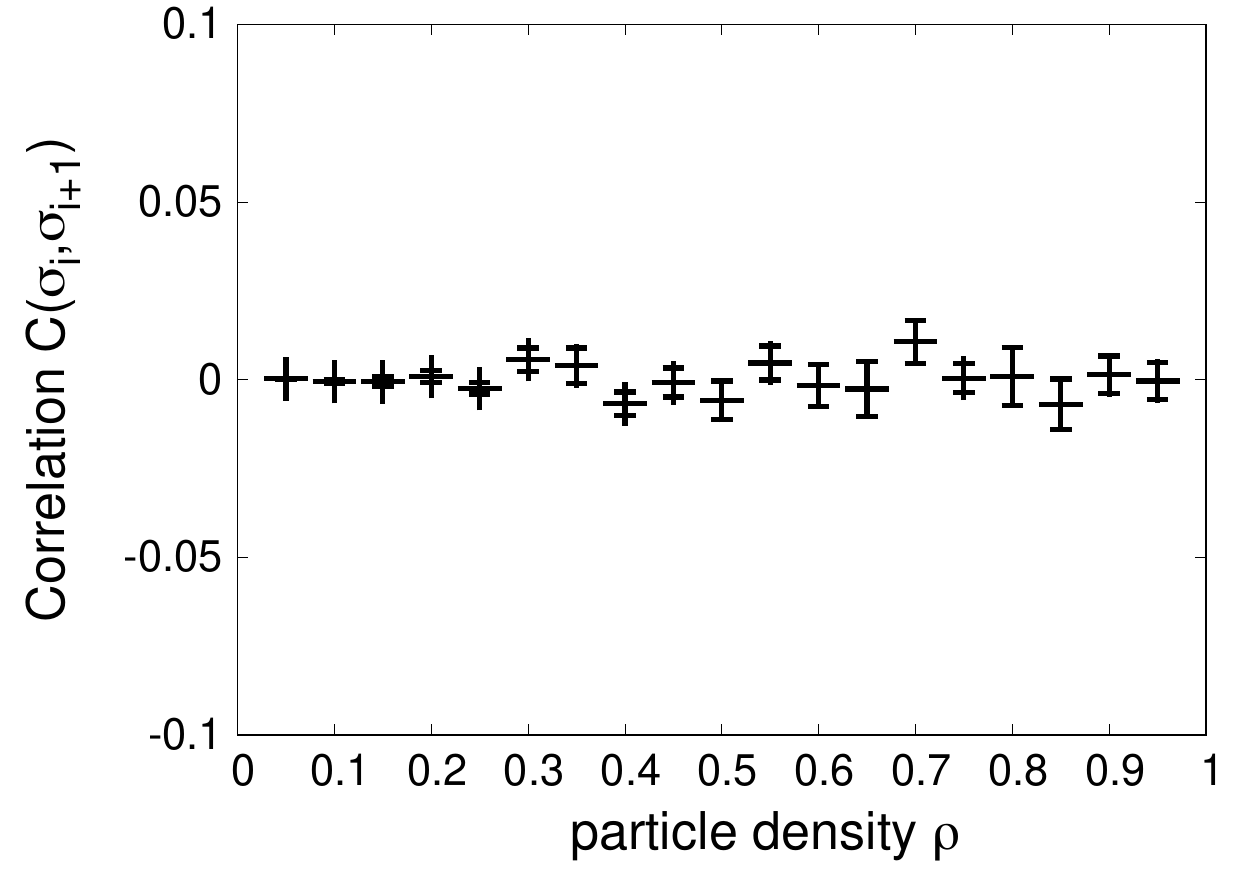}}
\subfloat[]{\includegraphics[width=0.48\columnwidth]{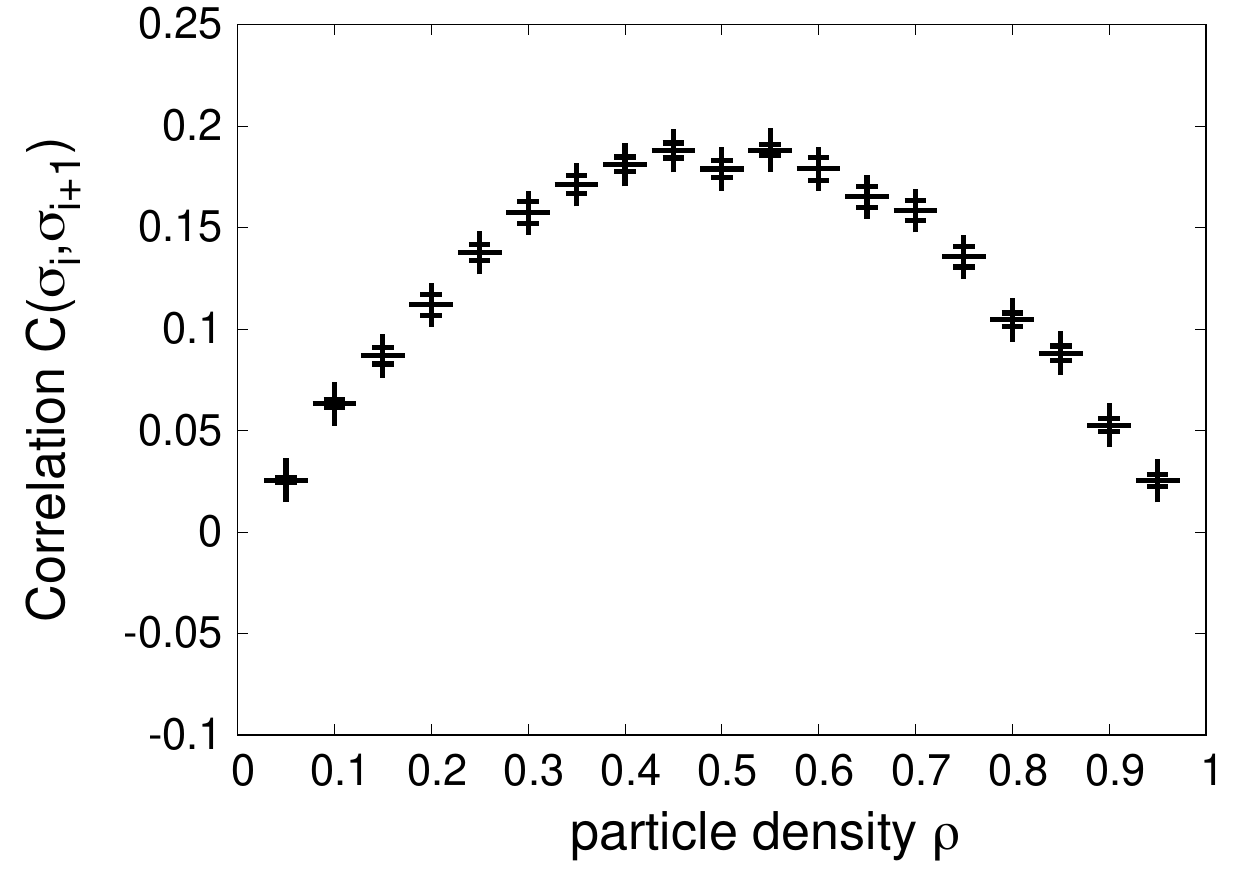}} \\
\subfloat[]{\includegraphics[width=0.48\columnwidth]{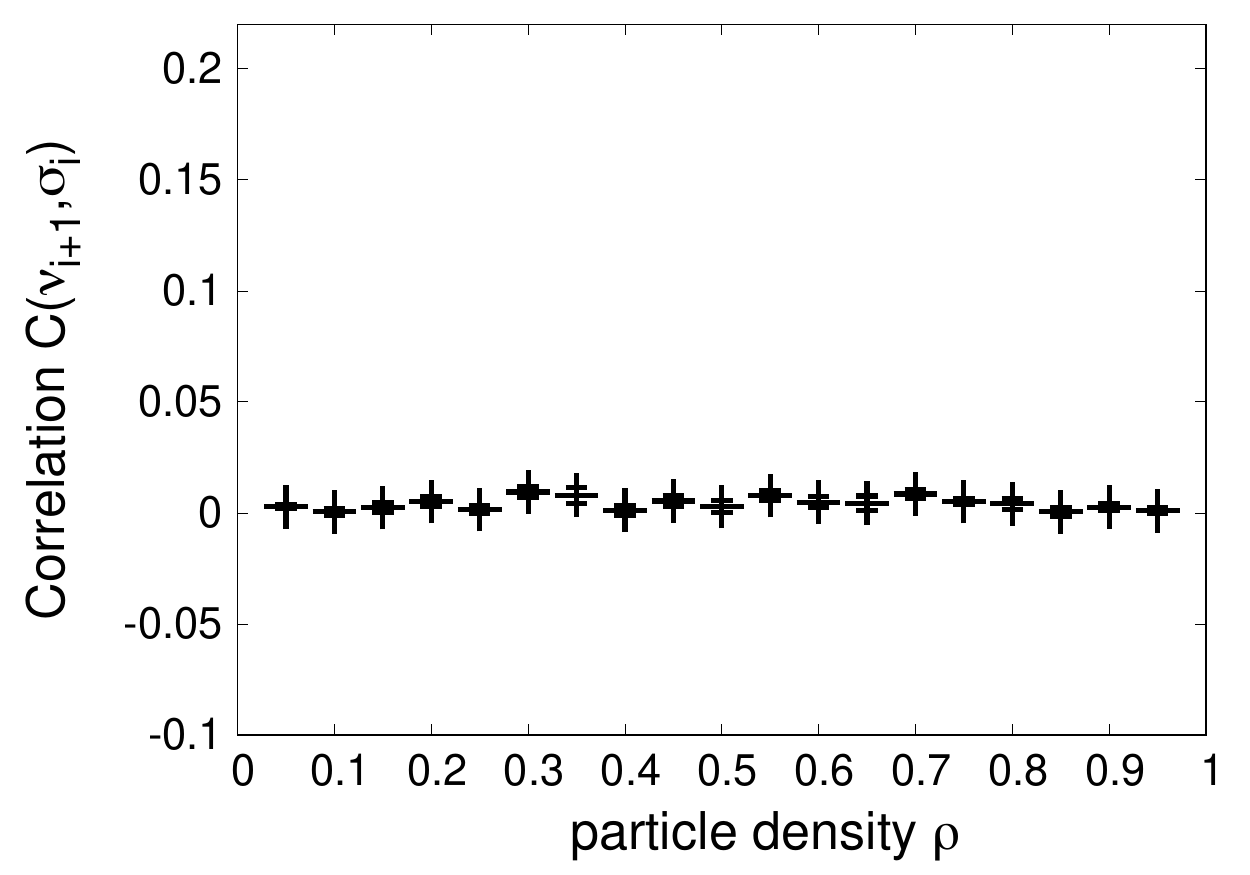}}
\subfloat[]{\includegraphics[width=0.48\columnwidth]{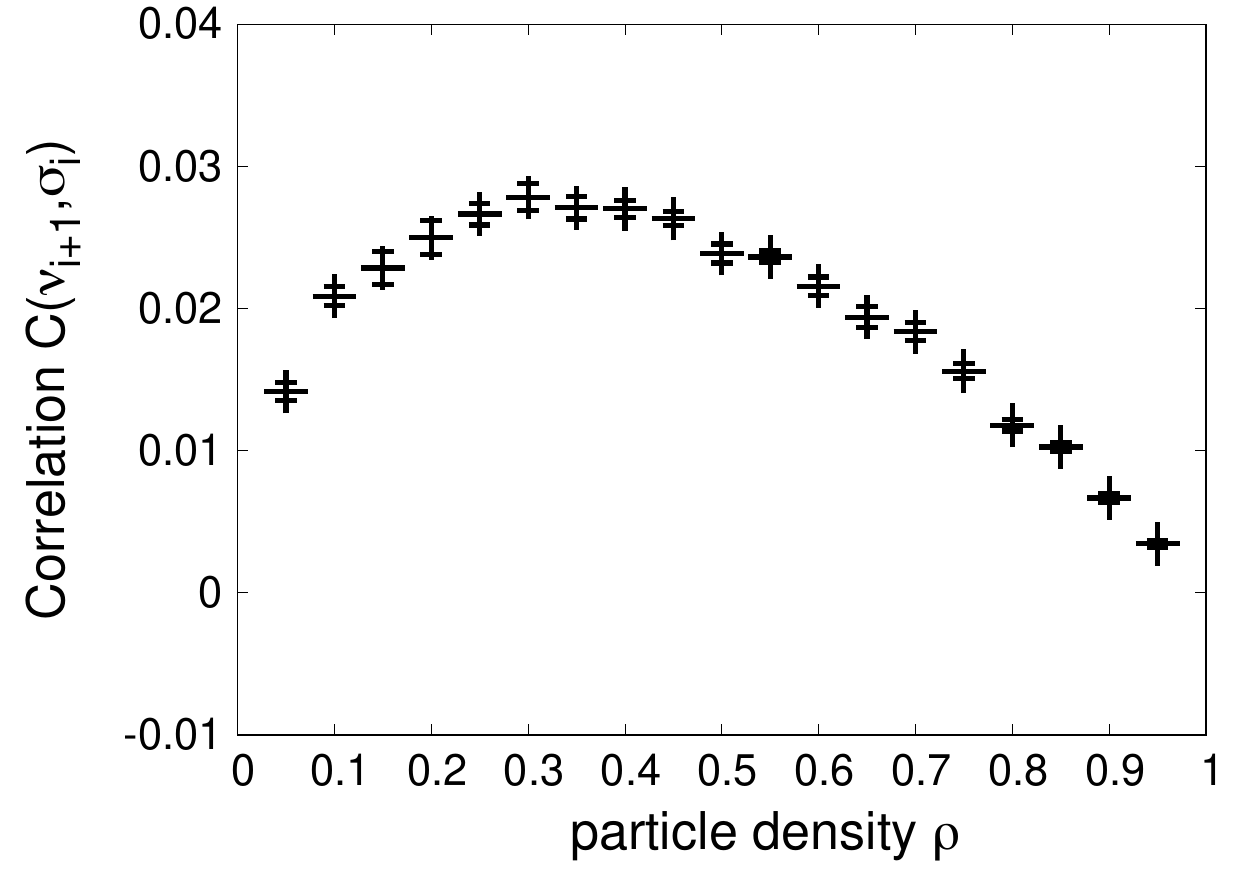}}
 \caption{\label{corr_nn_rho_fig} Correlations between neighboring sites. Upper row: $C(\sigma_{i},\sigma_{i+1}) = \langle \sigma_{i} \sigma_{i+1} \rangle - \rho^{2}$ as function of $\rho$, for $T = 100000/p$ and $L = 1000$ and (a) $k_- = 5p,\, k_+ = 5p$. (b) $k_- = 0.01p,\, k_+ = 0.002p$. Lower row: $C(\nu_{i+1},\sigma_{i}) = \langle \nu_{i+1} \sigma_{i} \rangle - \rho \rho_{d}$ as function of $\rho$, for $T=100000/p$ and $L=1000$ and (c) $k_- = 5p,\, k_+ = 5p$. (d) $k_- = 0.01p,\, k_+ = 0.002p$.}
 \end{figure}
 


To see if this relationship can be derived analytically, we consider a simple mean-field theory. Since for $p_d=0$ we have $p_{i} = p (1-\nu_{i+1})$, 
we have 
\bq
	J_{i} = p\langle (1-\nu_{i+1}) \sigma_{i}(1-\sigma_{i+1}) \rangle.
\eq
If we neglect correlations in the random variables $\nu_{i+1}$ and $\sigma_i$, we obtain $J_{i} \approx p \langle 1-\nu_{i+1} \rangle \langle \sigma_{i} \rangle \langle \sigma_{i+1} \rangle$. In the steady state, we have $J_{i} = J$, $\langle \sigma_{i} \rangle = \langle \sigma_{i+1} \rangle = \rho$ and $\langle \nu_{i} \rangle = k_+/(k_+ + k_-) =: \rho_{d}$ (the ``defect density'') \footnote{This follows directly from the equilibrium defect binding equation, $k_-\, \rho_d = k_+ (1 - \rho_d)$ (Langmuir kinetics)}, so that we get the \emph{naive} mean field approximation
\begin{align}
\label{naive_mf_eq}
J \approx p  \, (1-\rho_{d}) \rho(1-\rho).
\end{align} 
We note that for $\rho_d=0$ we recover the standard TASEP's exact relationship, Eq. (\ref{eq:jvsrho_standard}). As we can see in Figs. \ref{J_rho_fig}(a,b) for large $k_-, k_+ \gtrsim p$ this naive mean field approximation is very accurate. We conclude that for fast defect turnover, sites behave effectively as having a static, effective hopping rate $p(1-\rho_d)$, similarly to what has been conjectured for localised, single dynamic defects \cite{turci_transport_2013}.

To see why this approach works so well for  $k_-, k_+ \gg p$, in Fig. \ref{corr_nn_rho_fig}(a) we plot the correlations $C(\sigma_{i},\sigma_{i+1})$ 
and $C(\nu_{i+1},\sigma_{i})$ (Eq. \ref{eq:corr}) as a function of the particle density $\rho$. The observed correlations are very weak, thus justifying our mean field approximation.


If $k_- \ll p$ and $k_+ \ll p$, however, this approximation breaks down, see Fig. \ref{J_rho_fig} (c,d) (blue line). Figure \ref{space-time_plot}(c,d) shows that the distribution of particles becomes inhomogeneous for $k_{+,-} \ll p$, with pronounced particle clusters and large gaps emerging. It is clear that in this case there will be correlations between particles at neighbouring sites, which is indeed what Fig. \ref{corr_nn_rho_fig}(b) shows. We also expect a correlation between particle occupation and defect occupation on its right neighbour site, $C(\sigma_i,\nu_{i+1})$, because a defect causes particles to pile up in front of the defect. Figure \ref{corr_nn_rho_fig}(d) indeed shows strong correlations. This is in contrast to the lack of correlations for fast defect dynamics (Fig. \ref{corr_nn_rho_fig}(c)).

To obtain an \emph{enhanced} mean field theory, which takes into account correlations between $\nu_{i+1}$ and $\sigma_{i}$, we make the approximation 
\bq
\label{mf_enhanced_eq1}
J \approx p\langle (1-\nu_{i+1})\sigma_{i}\rangle \langle (1-\sigma_{i+1}) \rangle \,\,\, .
\eq
The first factor in this expression, $p\langle (1-\nu_{i+1})\sigma_{i}\rangle =: \bar \tau_f^{-1} \rho$, corresponds to the current of particles if there were no particle exclusion. Here, $\bar \tau_f$ is the \emph{free} mean waiting time (according to Eq. (\ref{J_i_def_eq})), in absence of exclusion interaction, and takes into account the correlation between a particle on site $i$ and defect occupation on site $i+1$. With this, the full current can be approximated as
\bq
J \approx \bar \tau_f^{-1}\rho(1-\rho).
\eq
If $k_+ \ll p$ we can neglect the rebinding of a defect after it has unbound. When a particle encounters a defect (probability $\rho_d$), it waits until the defect unbinds (rate $k_-$), and then it hops with rate $p$. When a particle encounters a no-defect site (probability $1-\rho_d$), it hops with rate $p$. Taking these two processes together, the waiting time of a particle can be approximated by
\bq
\label{tau_eq}
\bar \tau_f \approx \rho_{d}\left(\frac{1}{k_-} + \frac{1}{p}\right) + (1 - \rho_{d}) \frac{1}{p} = \frac{\rho_{d}\frac{p}{k_-} + 1}{p} .
\eq
The steady-state current is thus
\begin{align}
\label{enhanced_mf_eq}
J \approx \bar \tau_f^{-1}\rho(1-\rho) \approx \frac{p}{1 + \frac{p}{k_-}\rho_{d}} \,\rho(1-\rho).
\end{align}
Figure \ref{J_rho_fig}(c,d) shows that this approximation is substantially more accurate than Eq. (\ref{naive_mf_eq}) in the limit of slow defect dynamics, $k_{+,-} \ll p$. The remaining discrepancy between Eq. (\ref{enhanced_mf_eq}) and simulations is due to the correlation between particle occupancies on neighbouring sites, $C(\sigma_{i},\sigma_{i+1})$. We note that this enhanced mean field theory is not valid for large $k_{+} \sim p$ since in that case rebinding of obstacles, before a particle can hop, cannot be neglected, and thus the approximation made in Eq. (\ref{tau_eq}) does not apply.

\begin{figure}
\subfloat[]{\includegraphics[width=0.48\columnwidth]{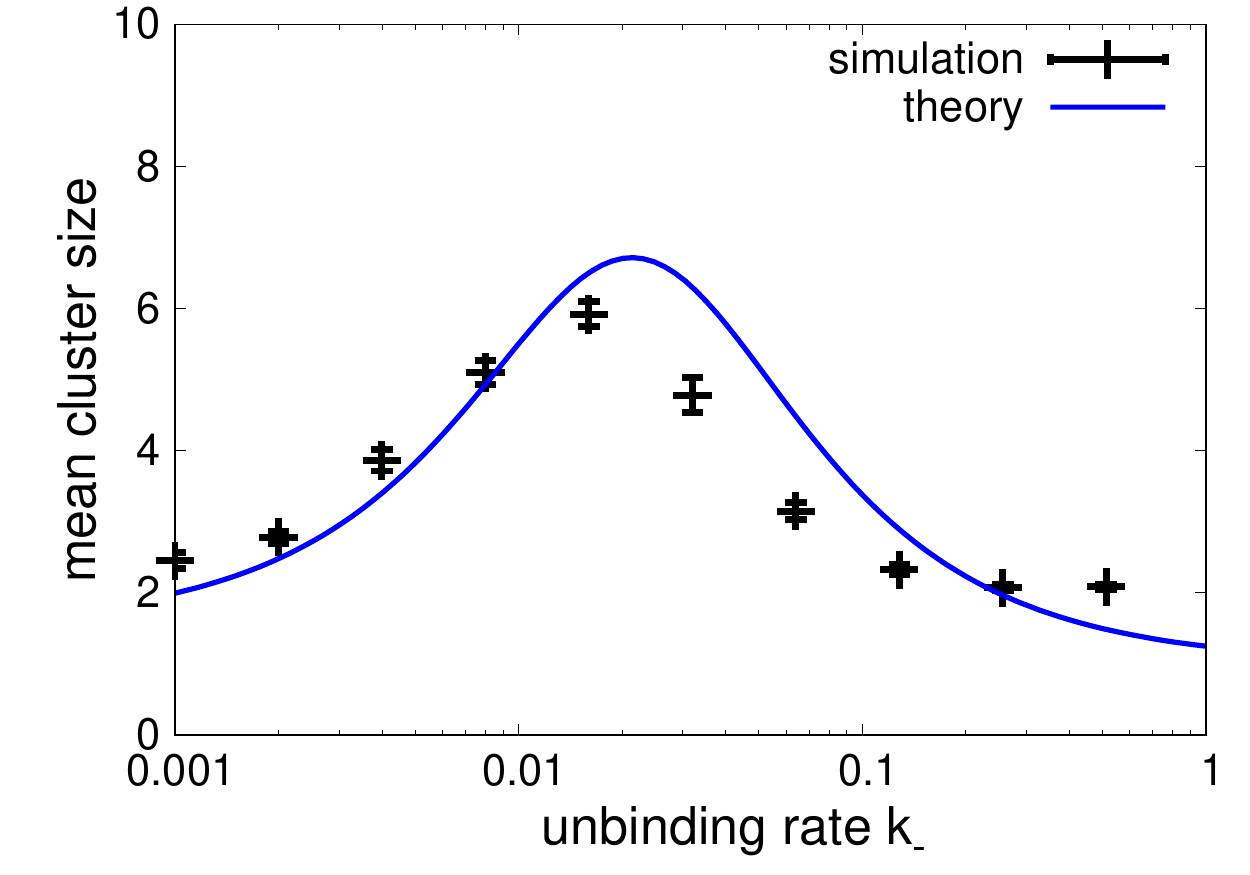}}
\subfloat[]{\includegraphics[width=0.48\columnwidth]{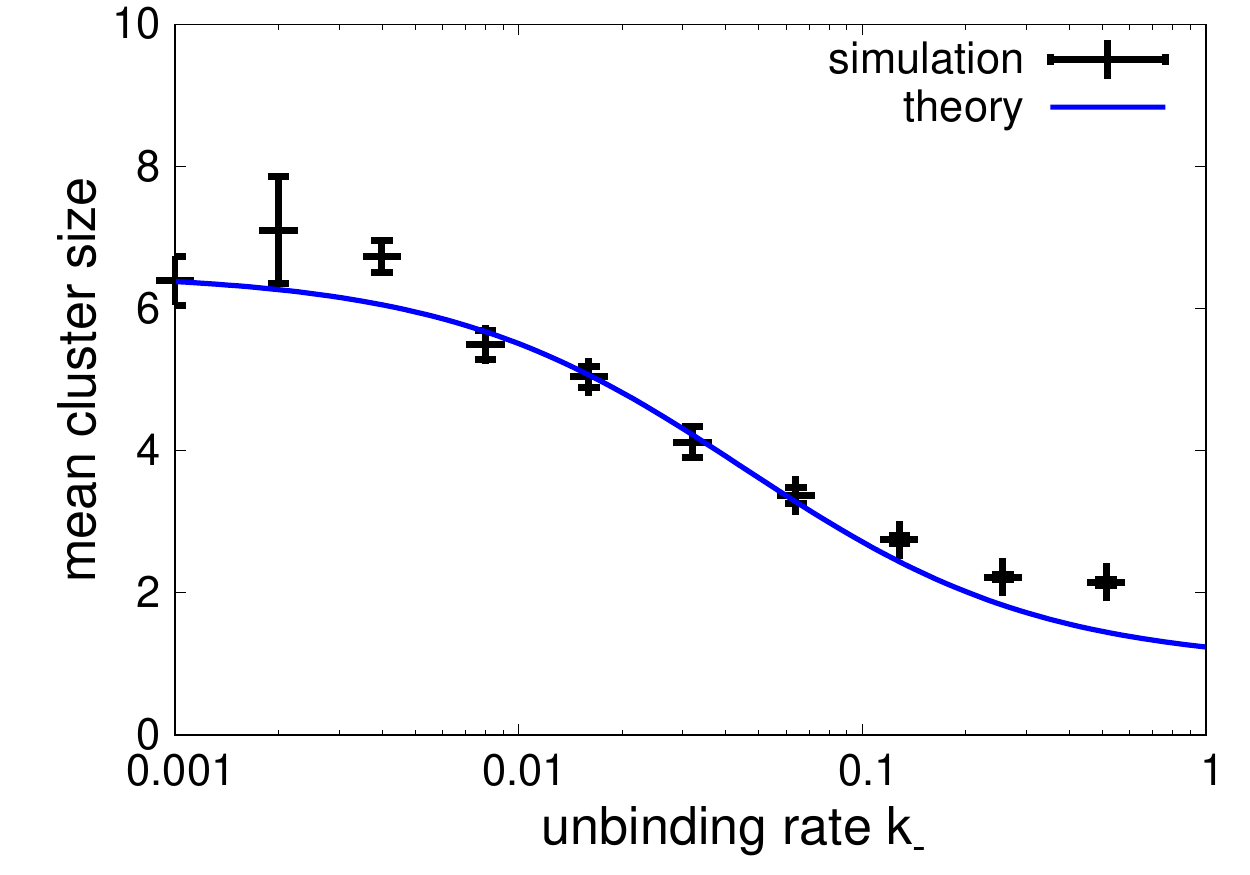}}
\caption{\label{av_cluster_fig} Mean cluster size (consecutive stretches of more than one particle), and analytical approximation, Eq. (\ref{mean_cluster_eq}) (blue line), for $T = 1000000/p$, $\rho=0.1$, $L=500$. (a) as a function of unbinding rate $k_-$ for fixed $k_+ = 0.001p$. (b) as a function of $k_-$ while $k_+ = 0.1 k_-$ is scaled so that $\rho_{d}$ remains constant.}
\end{figure}

To understand the observed inhomogeneity of the particle distribution, we measure the mean cluster size while varying the defect binding/unbinding rates $k_+,k_-$. A cluster is defined as a non-interrupted stretch of particles (more than one particle). Figure \ref{av_cluster_fig}(a) shows the mean cluster size as a function of $k_-$, for fixed $k_+ = 0.001p$. We observe a non-monotonic dependence of the cluster size on $k_-$, with the maximum size at $k_- \approx 0.02p$. In Fig. \ref{av_cluster_fig}(b), on the other hand, the mean cluster size is scaled with both $k_-$ and $k_+ = 0.1 k_-$ while keeping the defect density $\rho_{d} = k_+/(k_+ + k_-) = 1/11$ fixed ($k_-$ shown on x-axis). In this case we do not see any significant peak (within error margins) in the mean cluster size.

We can estimate a typical cluster size as follows. A particle cluster is initiated at site $i$ when a particle is blocked by a defect site, $\nu_{i+1} = 1$, while trailing particles between that defect and the following defect towards the left -- let it reside at site $(i-d)$ -- pile up. We first consider what happens in two limiting cases when $k_-$ is either very small or very large. For sufficiently small $k_-$, defect unbinding can be neglected and all particles residing between $i-d$ and $i$ pile up to form a cluster. In total, there are on average $\rho_c \, d$ particles between two neighbouring defects, where $\rho_c$ is the local particle density, which may differ from the mean density of particles $\rho$ (we shall argue below that $\rho_c \approx 0.5$). In that limiting case, the mean final cluster size is $\bar l \approx \rho_c \, \langle d \rangle = \frac{\rho_c}{\rho_d}$. If $k_-$ is large, however, such that the defect at site $i$ unbinds before all particles between sites $i-d$ and $i$ pile up completely, the growth of clusters is limited by $k_-$. In that case particles flow into the cluster with free current $J_f = p \rho_c(1 - \rho_c)$ (since there are no other defects between two neighbouring ones), for a mean time $t_u = 1/k_-$, which leads to a mean final cluster size $\bar l \sim J/k_- = \frac{p}{k_-}\rho_c(1-\rho_c)$ for $k_- \to \infty$. We note that the final cluster size is an over-estimation in this limit, since at the time point of defect unbinding, it exists only for a short time (in contrast to the case of small $k_-$, when a saturated cluster can exist for a long time). However, the final cluster size yields the correct magnitude of clusters, and allows to compare and interpolate the two limiting cases, as follows.

To determine the cluster dynamics for intermediate time scales $k_-$ we consider the cluster growth dynamics in more detail. Particles between two defects accumulate with rate $J_f \, t$, where $t$ is the time after cluster initiation. The cluster growth can be stopped by the two events, that (i) the initiating defect at site $i$ {\bf u}nbinds, with rate $k_-$, which is related to the time scale $\bar t_{u} = 1/k_-$, or (ii) all particles between the defects at site $i-d$ and $i$ (on average $\rho_c/\rho_d$ particles) have been {\bf e}xhausted, which happens when the cluster has grown to include all particles between the defects, after a time scale $t_e$ defined by $J_f \, \bar t_{e} = \rho_{c}/\rho_d$, thus $\bar t_{e}= 1/(p\rho_{d}(1-\rho_{c}))$. The stochastic events (i) and (ii) occur independently from each other. The probability that the cluster is still growing by time $t_c$ can then be approximated as $P(t_{c}) = P(t_{u}) P(t_{e}) \sim e^{-t/t_{u}} e^{-t/t_{e}} = e^{-(t_{u}^{-1} + t_{e}^{-1}) t}$. The mean time of cluster growth is thus $\bar t_{c} = (t_{u}^{-1} + t_{e}^{-1})$ and hence the mean cluster size is
\begin{align}
\label{mean_cluster_eq}
\bar l \approx J_f \bar t_{c} = \frac{p\rho_{c}(1-\rho_{c})}{k_- + p(1-\rho_{c})\rho_{d}}.
\end{align}
This indeed interpolates between the limiting cases for small $k_- \ll p(1-\rho_c)\rho_d$ and large $k_- \gg p(1-\rho_c)\rho_d$, as discussed above.

Finally, we note that the density $\rho_c$ corresponds to the density of particles flowing out of a dissolving cluster, later initiating a new one. It was argued in Ref. \cite{turci_transport_2013} that dissolving clusters behave locally like a maximum current phase in absence of defects, such that this density corresponds to the maximum current density of the TASEP, $\rho_c \approx 0.5$.

Inserting $\rho_{c} =0.5$ and $\rho_{d} = k_+/(k_+ + k_{-})$ into Eq. (\ref{mean_cluster_eq}) we obtain the mean cluster size $\bar l$ as a function of $k_-$ and $k_+$. Figure \ref{av_cluster_fig}(a) shows the mean cluster size $\bar{l}$ as a function of $k_-$ for fixed $k_+ = 0.001p$ while Fig. \ref{av_cluster_fig}(b) shows $\bar{l}(k_-)$ for $k_+ = 0.1 \, k_-$ so that $\rho_{d} = 1/11$ is kept fixed. The blue line shows the result from Eq. (\ref{mean_cluster_eq}), which confirms that our approximate calculation correctly estimates mean cluster size for both cases. As expected, for large $k_-$ our theory, which considers clusters at their maximum size, over-estimates the simulated value. Crucially, however, the peak in $\bar{l}(k_-)$ in the case of fixed $k_+$ is accurately reproduced by Eq. (\ref{mean_cluster_eq}). This peak is due to the competing limiting cases: For low $k_-$, as long as $k_-$ is much smaller than $p(1-\rho_{c})\rho_{d}$, $\bar l \approx \rho_c/\rho_d \propto (k_- + k_+)/k_+$, which increases with $k_-$. Beyond this point the blocking defect (on average) unbinds before the cluster can grow to its full size (event (i)), and the cluster size is determined by the defect life time, which is $\sim 1/k_-$. Thus, in this regime the mean cluster size decreases with $k_-$.

\subsection{Constrained defect dynamics and $p_d = 0$}
\label{constrained_sec}

\begin{figure*}[h]
\subfloat[]{\includegraphics[width=0.48\columnwidth]{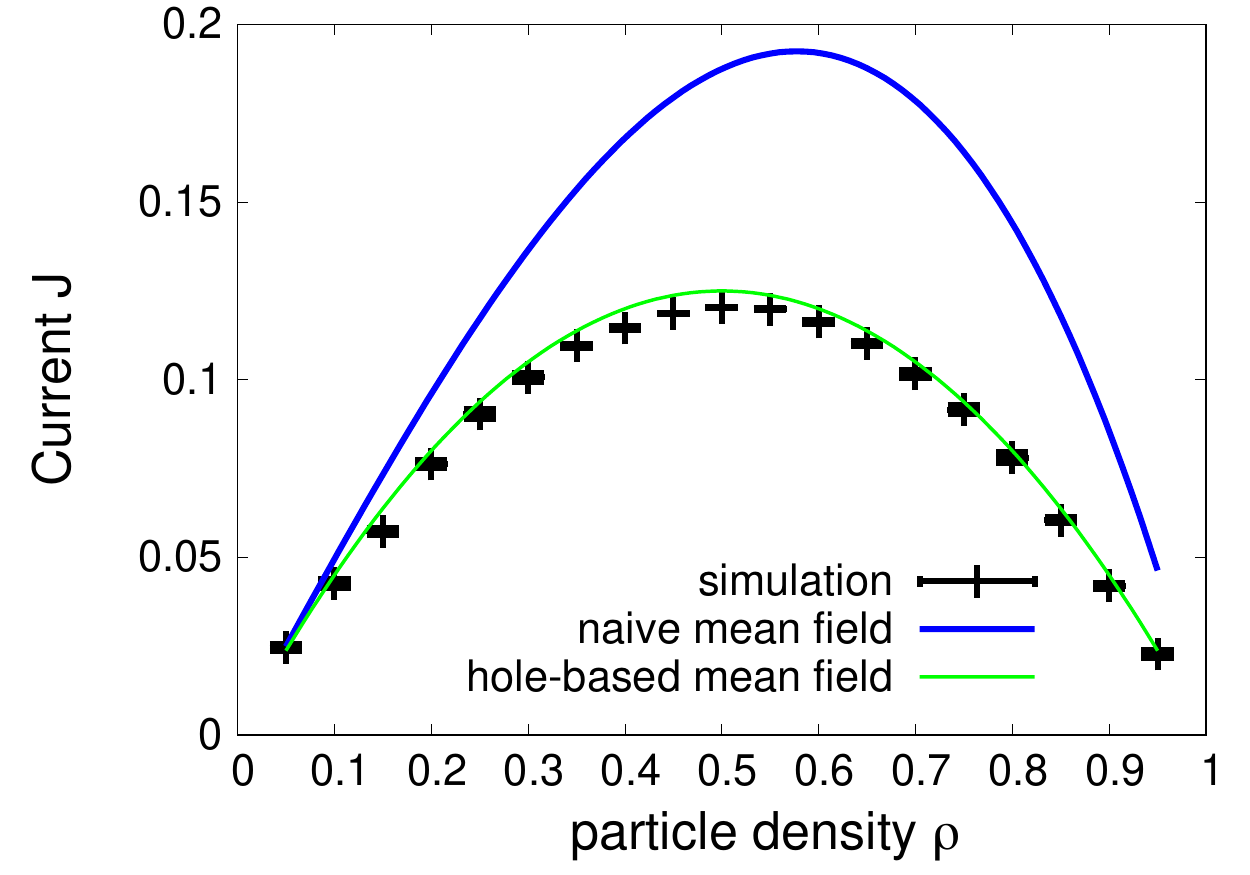}}
\subfloat[]{\includegraphics[width=0.48\columnwidth]{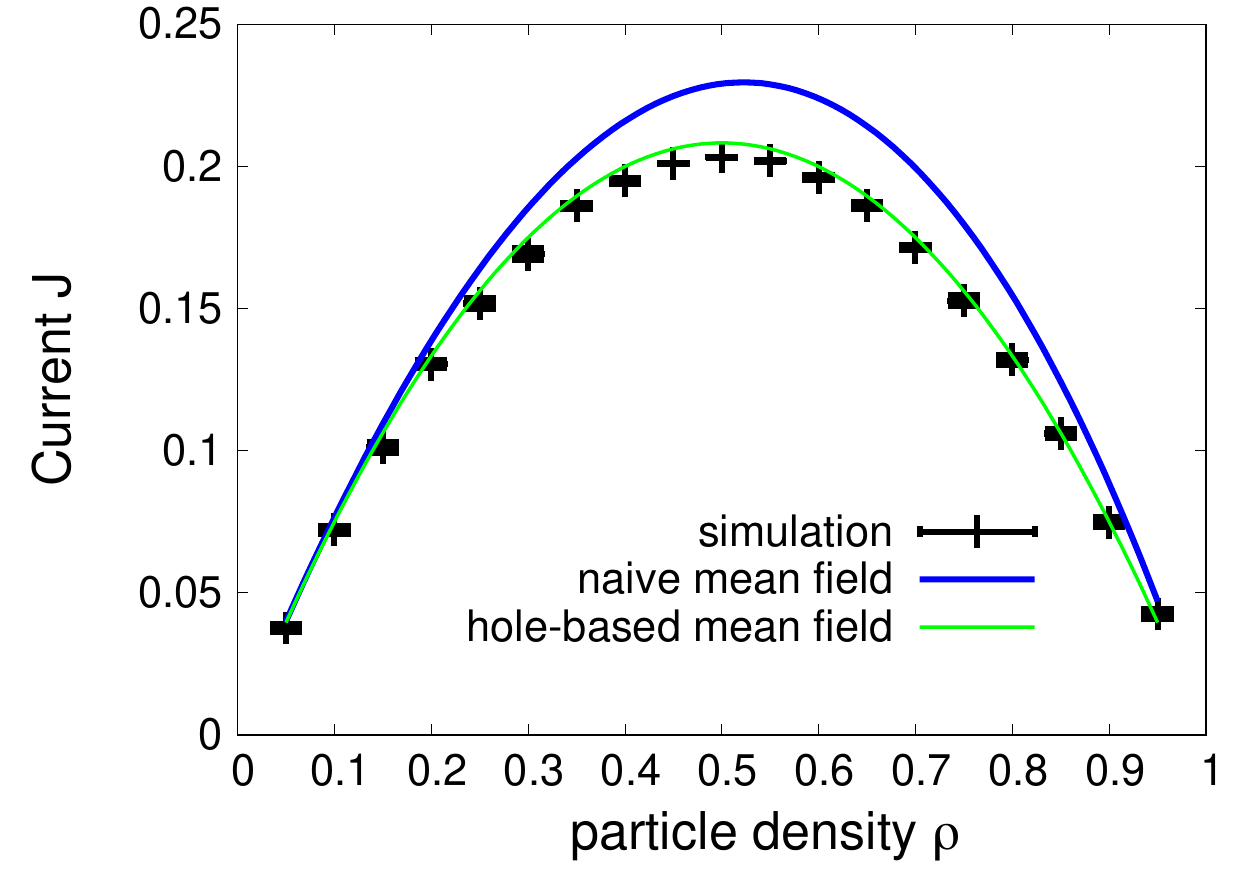}} \\
\subfloat[]{\includegraphics[width=0.48\columnwidth]{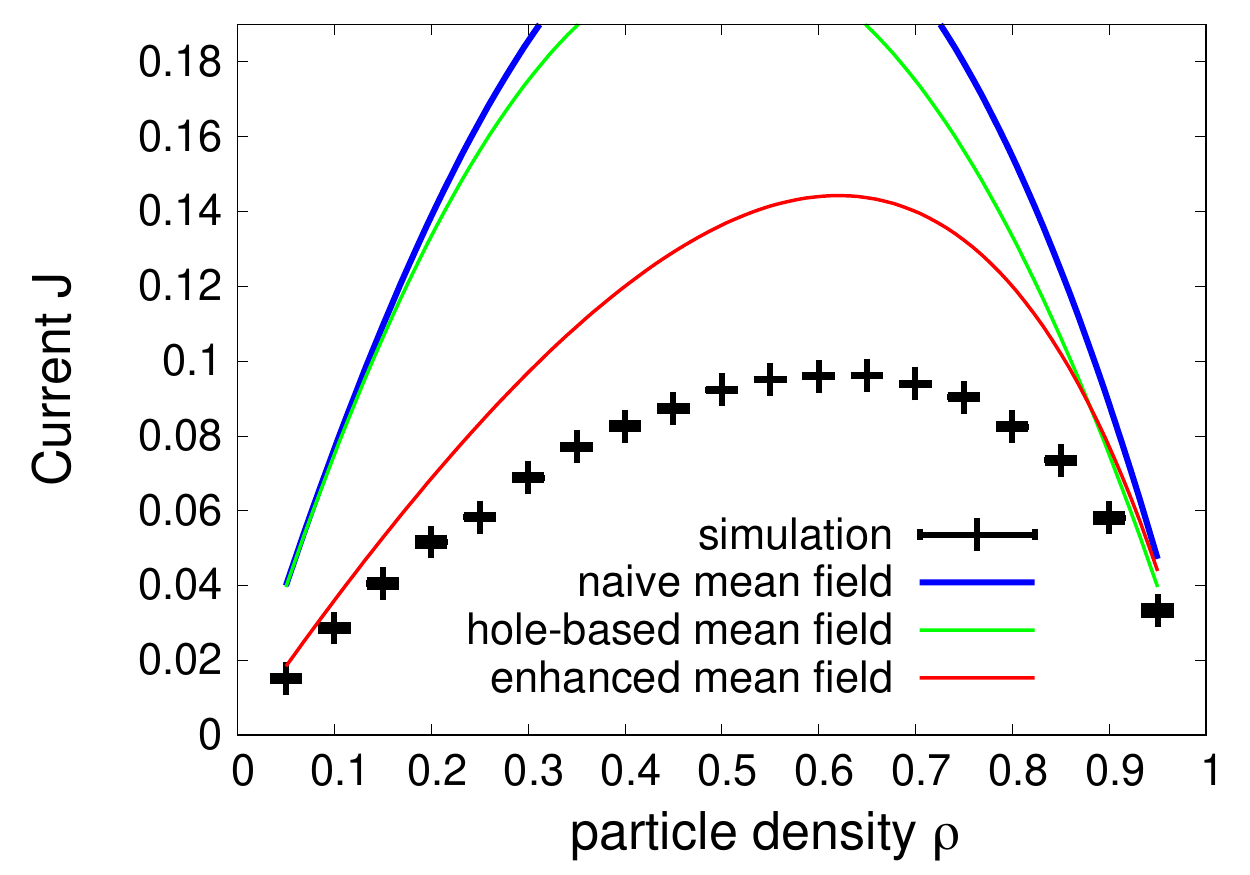}}
\subfloat[]{\includegraphics[width=0.48\columnwidth]{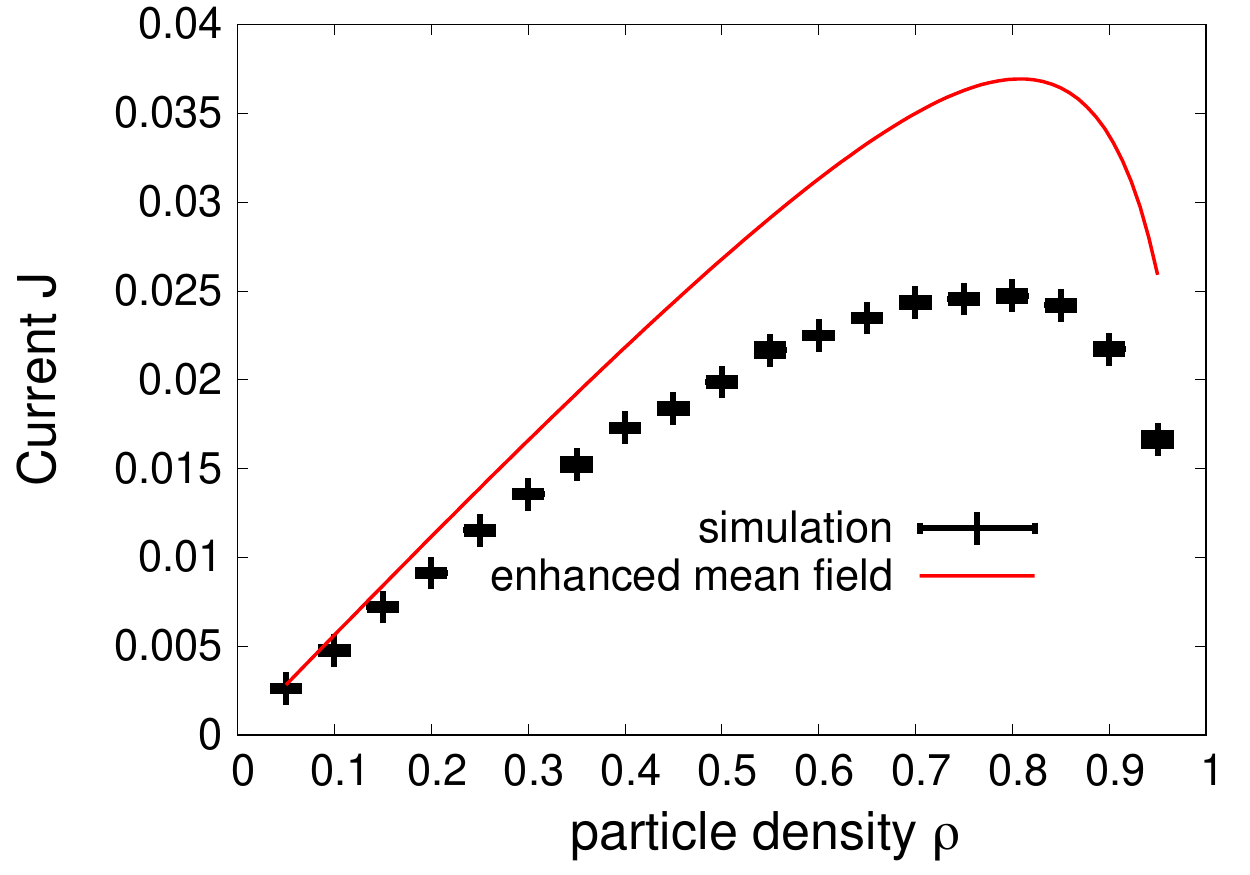}} \\
\caption{\label{J_rho_restr_fig} Particle current $J$, in units of $p$, as a function of the particle density $\rho = \langle \sigma_{i} \rangle$ for constrained defect binding (Eq. (\ref{model_varB})), $T=100000/p$ and $L=1000$ and different rates of defect binding/unbinding rates $k_+,k_-$. The blue line is the naive mean field approximation, Eq. (\ref{naive_mf_constr}), the green line is the hole-based mean field approximation, Eq. (\ref{hole-based_MF_eq2}), and the red line is the enhanced mean field approximation, Eq. (\ref{mf_enhanced_constr_eq}). (a) $k_- = 5p,\, k_+ = 5p$. (b) $k_- = 5p,\, k_+ = p$. (c) $k_- = 0.1p,\, k_+ = 0.02p$. (d) $k_- = 0.01p,\, k_+ = 0.002p$.}
\end{figure*}

\begin{figure*}[h]
\subfloat[]{\includegraphics[width=0.48\columnwidth]{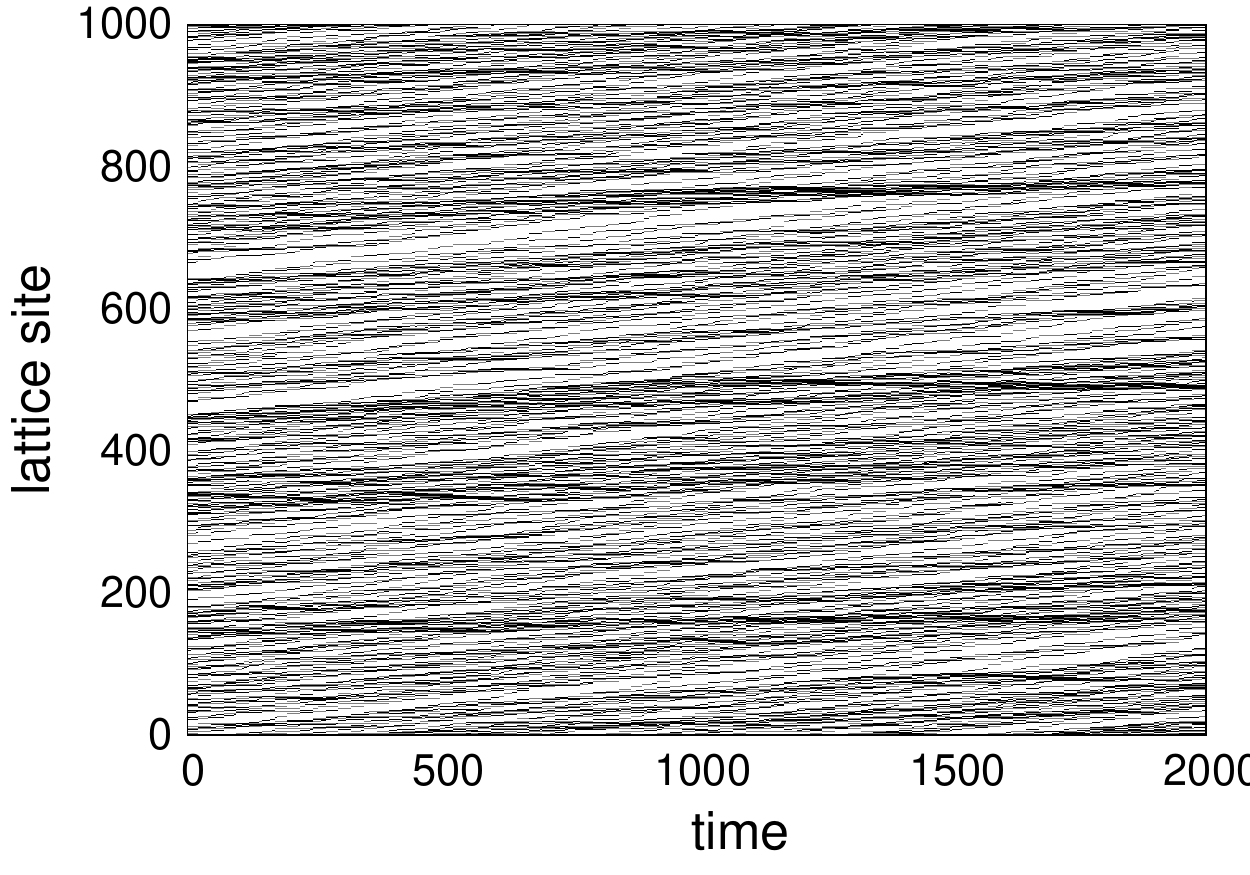}}
\subfloat[]{\includegraphics[width=0.48\columnwidth]{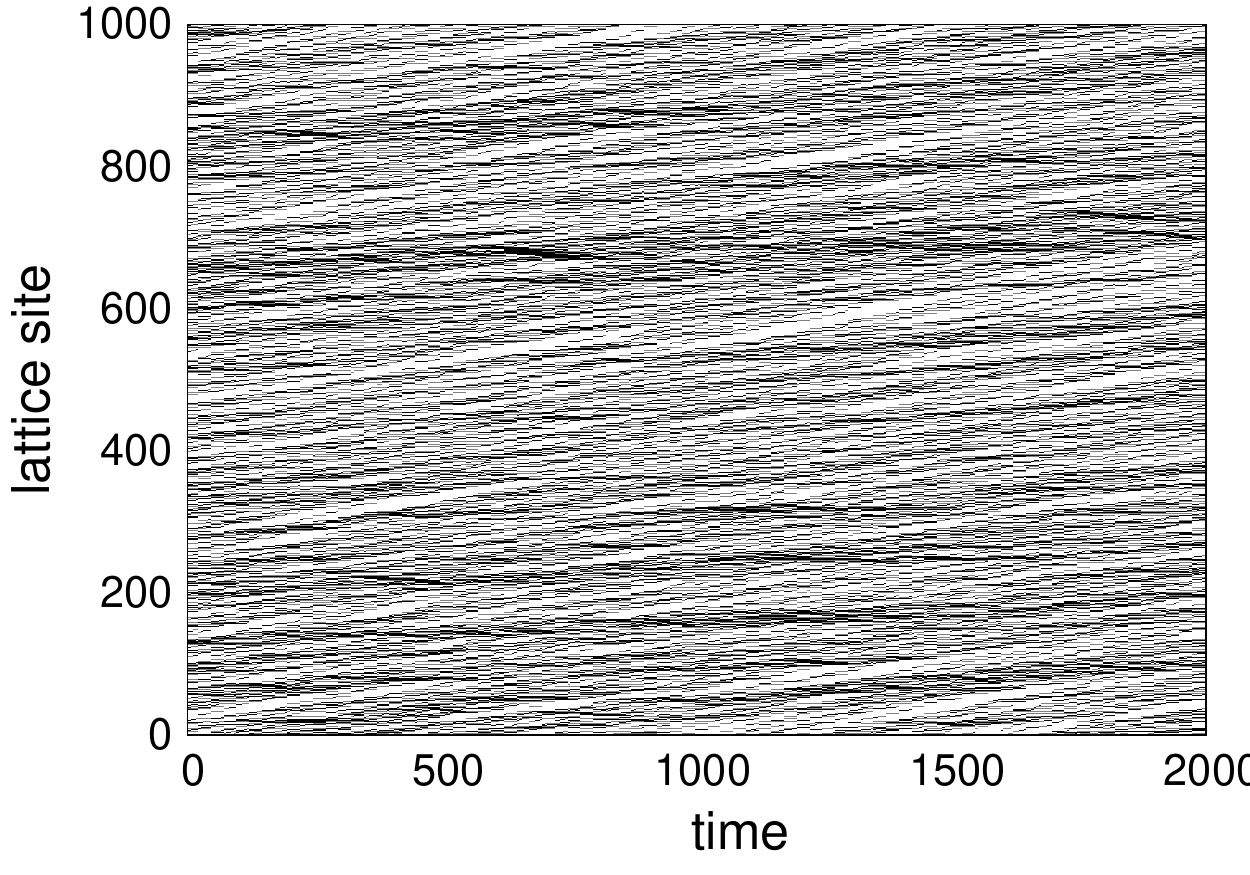}} \\
\subfloat[]{\includegraphics[width=0.48\columnwidth]{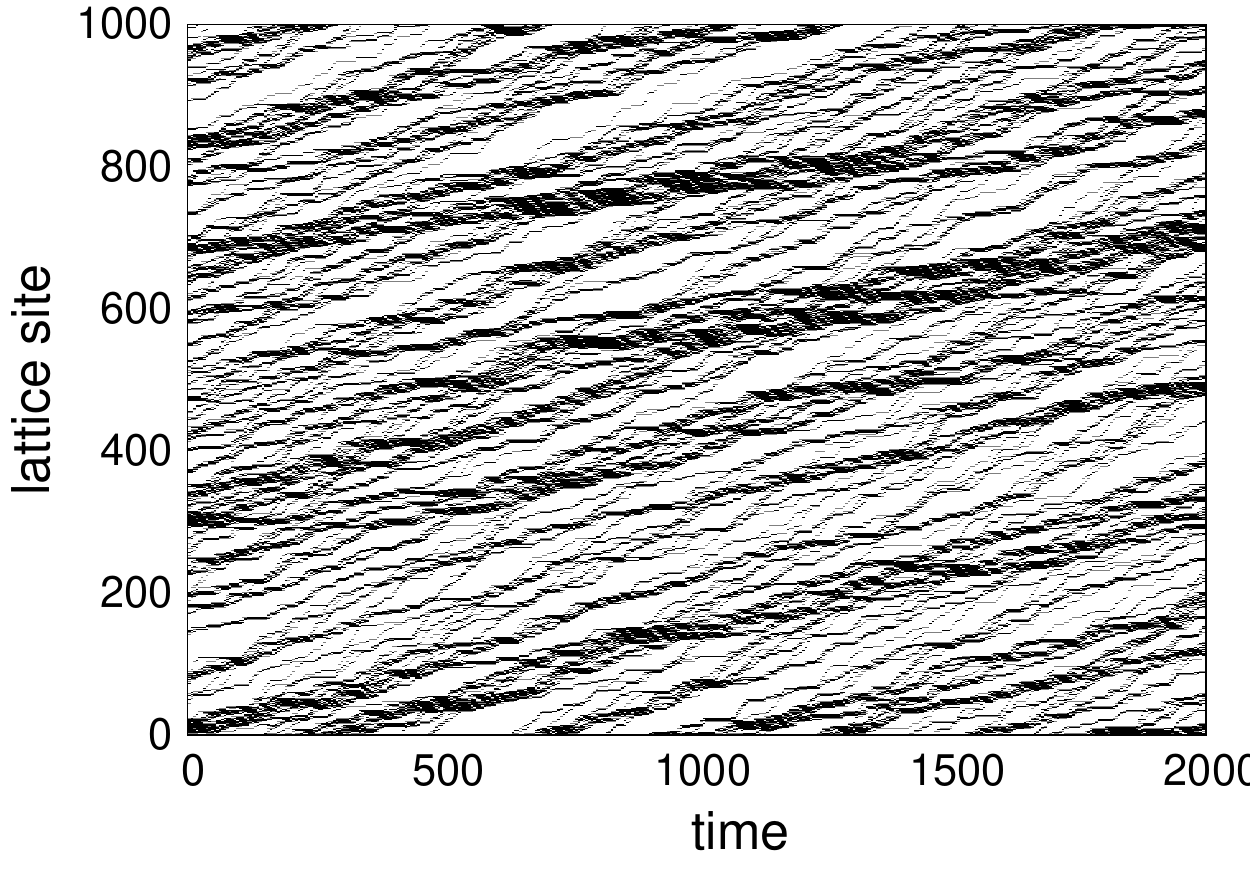}}
\subfloat[]{\includegraphics[width=0.48\columnwidth]{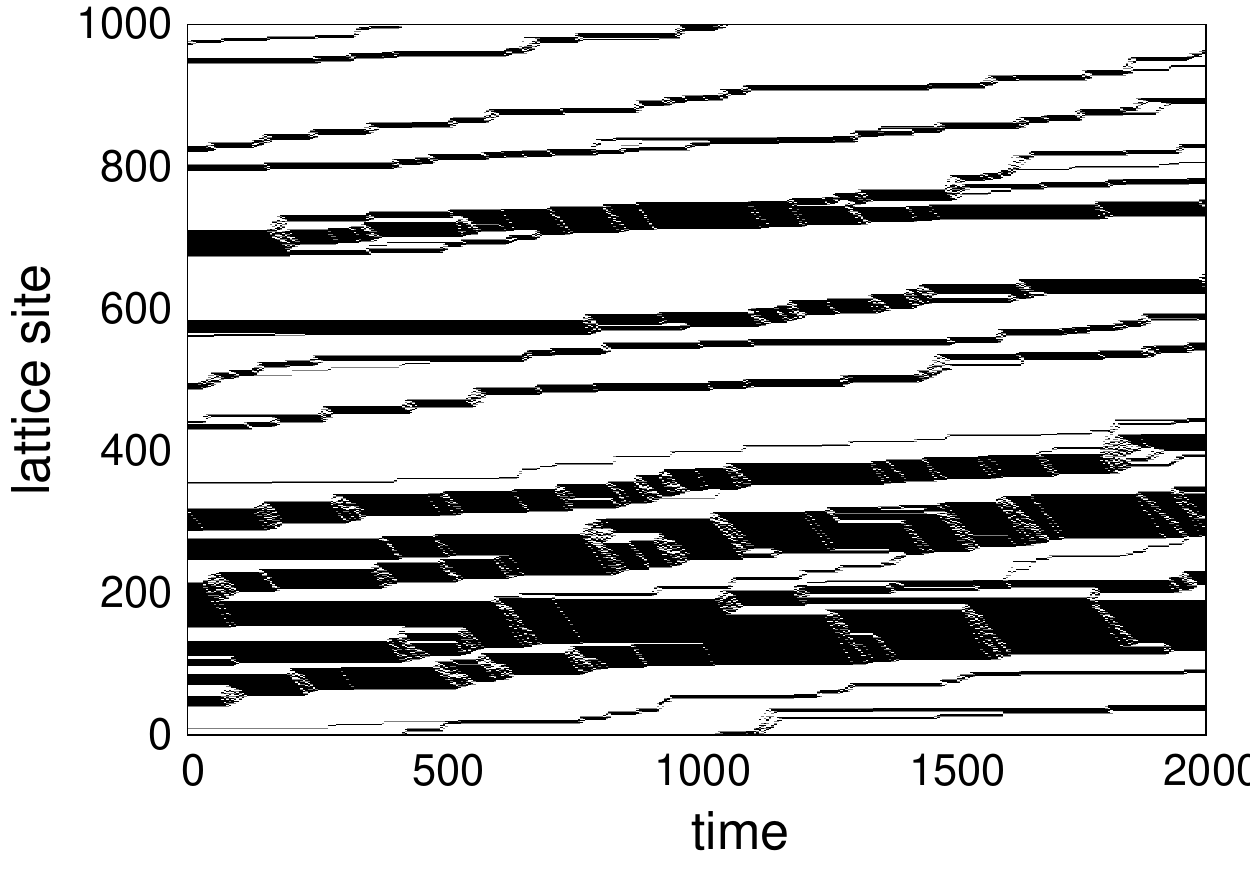}}
\caption{\label{space-time_restr_plot} Space-time plots for $\rho=0.3$ and constrained defect binding (Eq. (\ref{model_varB})); each pixel denotes a particle, where the $y-$axis denotes the lattice site and the $x-$axis time, in units of $p^{-1}$ .  (a) $k_- = 5p,\, k_+ = 5p$. (b) $k_- = 5p,\, k_+ = p$. (c) $k_- = 0.1p,\, k_+ = 0.02p$. (d) $k_- = 0.01p,\, k_+ = 0.002p$.}
\end{figure*}

Now we consider the constrained variant of the model in which obstacles can only bind if a site is not occupied by a particle (Eq. (\ref{model_varB})). Figure \ref{J_rho_restr_fig} shows $J(\rho)$ for the same parameters as in Fig. \ref{J_rho_fig}. The curve  $J(\rho)$ is approximately parabolic for fast defect turnover ($k_{+,-}\gtrsim p$), resembling Eq. (\ref{eq:jvsrho_standard}) for the ordinary TASEP. However, as $k_{+,-}$ decrease, $J(\rho)$ becomes skewed to the right, with the maximum shifted to $\rho_{\rm max} > 1/2$. Figure \ref{space-time_restr_plot} shows the corresponding space-time plots; particles become non-uniformly distributed and form clusters for low $k_{+,-}$, similarly as for the unconstrained binding model. 

In order to understand what aspects of particle-defect interactions are responsible for the skewed $J(\rho)$, we shall consider a range of mean-field approximations of increasing complexity. We start from the naive mean field approach which neglects any correlations. Similar to the unconstrained case, we have $J \approx p(1-\langle \nu_{i+1} \rangle)\rho(1-\rho)$, however, since the binding rate depends on the particle occupation, the mean field equilibrium defect density is $\rho^{(c)}_d(\rho) := \langle \nu_{i+1} \rangle \approx (1-\rho)k_+/(k_+ + k_-)$, i.e. it depends on the particle density\footnote{This follows again from the equilibrium binding equation, $k_- \rho^{(c)}_d = (1 - \rho - \rho^{(c)}_d) k_+$, where we used that defect- and particle occupation is mutually exclusive.}. We obtain 
\bq
\label{naive_mf_constr}
J(\rho) = p(1-\rho^{(c)}_d(\rho))\rho(1-\rho),
\eq
which predicts that the CDR should be skewed around $\rho=1/2$. Nonetheless, as shown in Figure \ref{J_rho_restr_fig}, this approximation is not appropriate for either value of the binding parameters $k_+,k_-$. In particular, for large $k_+,k_-$ the simulation data shows no significant skewness.  

To improve our approximation, we can focus on the dynamics of holes instead of particles. Defects can bind to holes (and only to holes) without constraints, thus this model variant corresponds to \emph{hole-wise} unconstrained dynamic disorder. Our \emph{hole-based} mean field approximation follows
\bq
\label{hole-based_MF_eq1}
J = \langle p (1 - \nu_{i+1}) (1-\sigma_{i+1}) \sigma_{i} \rangle \approx p \langle (1-\nu_{i+1}) (1-\sigma_{i+1}) \rangle \langle \sigma_{i} \rangle,
\eq
which is similar to the mean field approximation in Eq. (\ref{mf_enhanced_eq1}), but here we ``pair'' $\nu_{i+1}$ with $\sigma_{i+1}$. Crucially, since defect dynamics on holes are unconstrained and thus independent of the holes' dynamic history, the probability that there is a hole without a defect on site $i+1$, $\langle (1-\nu_{i+1}) (1-\sigma_{i+1}) \rangle$, is exactly $(1-\rho_d)(1-\rho)$, where $\rho_d = k_+/(k_+ + k_-)$ is the \emph{unconstrained} defect density. Thus, the hole-based mean field approximation for the current is
\bq
\label{hole-based_MF_eq2}
J(\rho) \approx p(1-\rho_d)\rho(1-\rho),
\eq
which is identical to the naive mean field theory of the unconstrained case, and which is indeed a symmetric CDR. In fact, we see that this hole-based mean field approximation matches well the CDR for fast defect turnover, Fig. \ref{J_rho_restr_fig}(a). This can also be understood intuitively: a particle can only hop if the next site is empty. Due to the fast equilibration of defects, the probability that a particle-free site is occupied by a defect is well approximated by the unconstrained equilibrium value $\rho_{d}$ and not by the real, constrained, defect density $\rho^{(c)}_d$.

\begin{figure}[h]
\subfloat[]{\includegraphics[width=0.48\columnwidth]{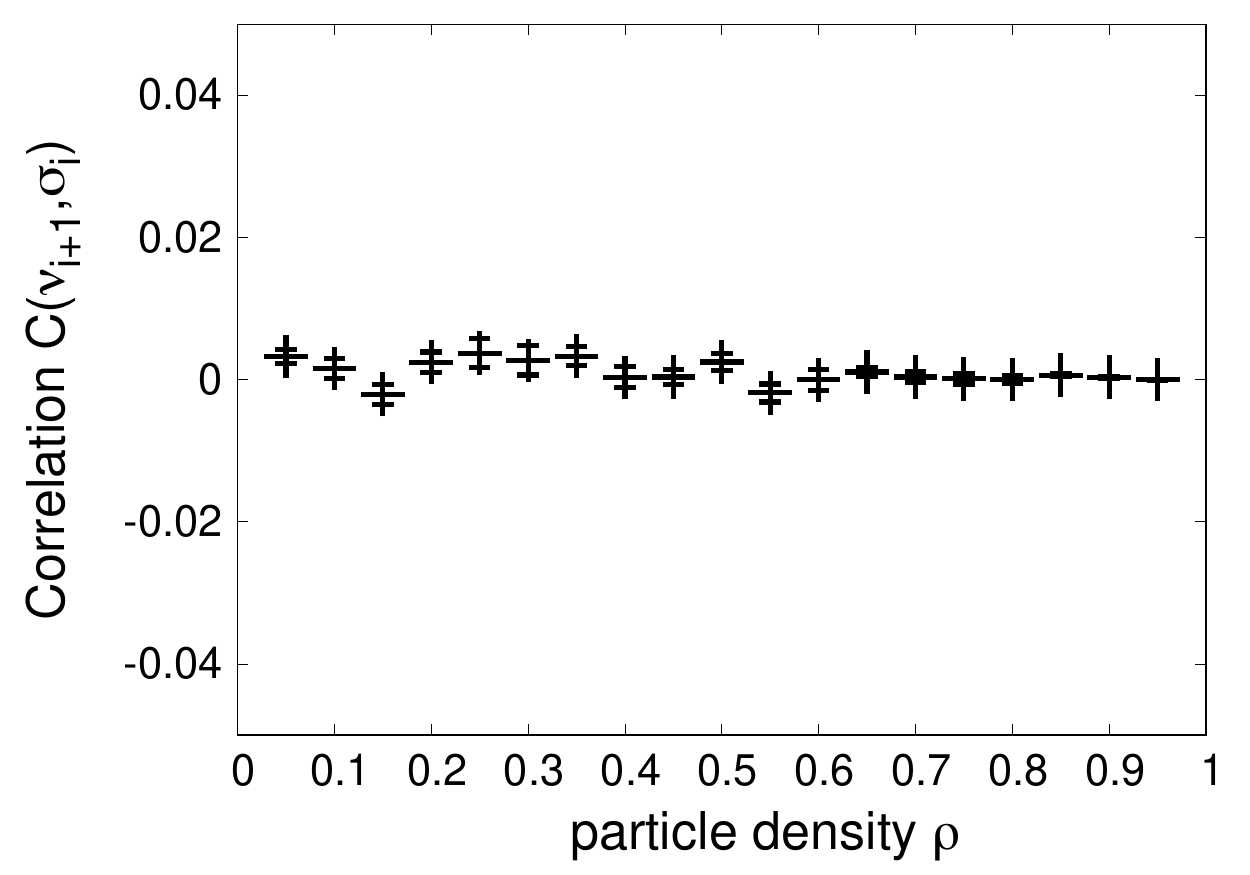}}
\subfloat[]{\includegraphics[width=0.48\columnwidth]{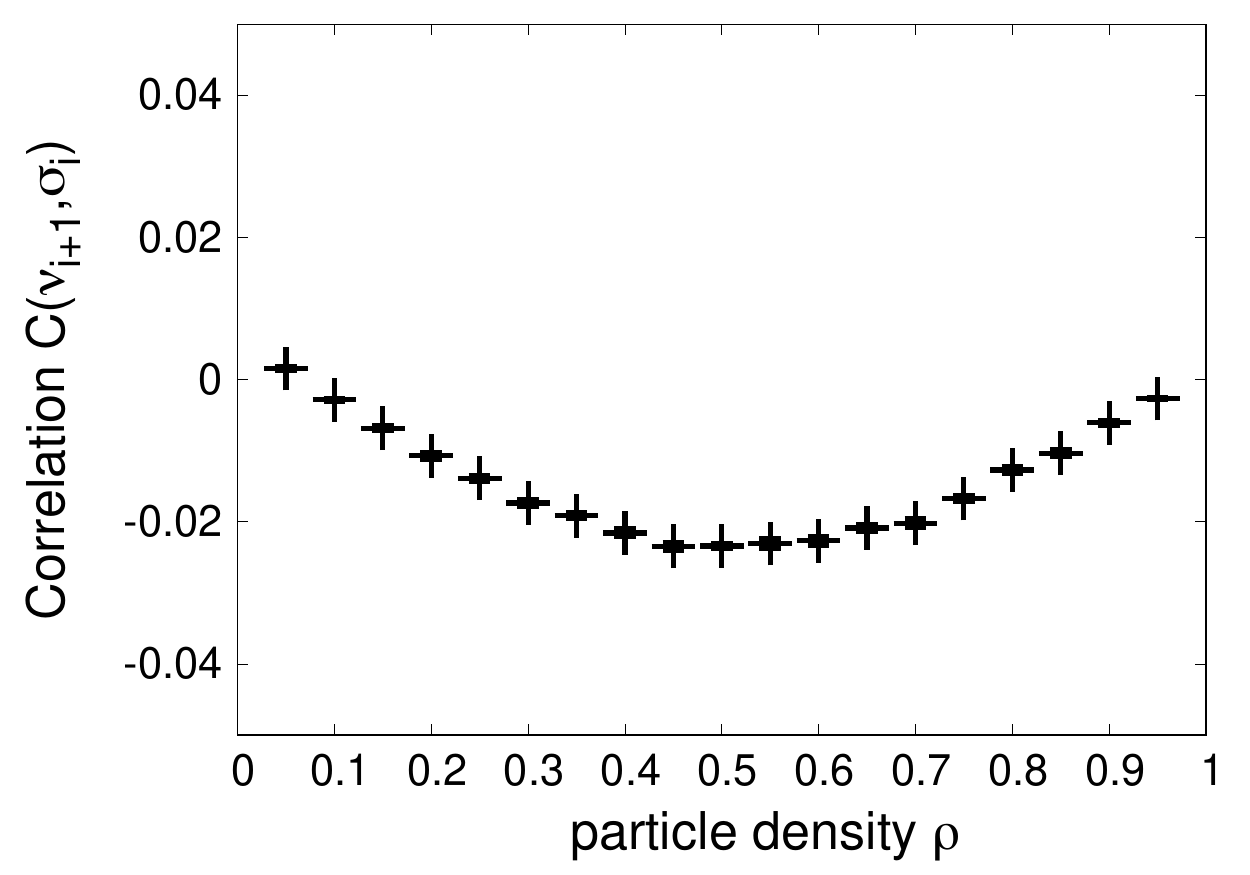}} \\
 \caption{\label{corr_nn_rho_restr_fig} Correlations between particle and defect occupation on neighbouring sites for constrained defect dynamics, $C(\sigma_{i},\nu_{i+1}) = \langle \sigma_{i} \nu_{i+1} \rangle - \rho\rho_{d}$ as function of $\rho$, for $T=100000/p$ and $L=1000$, and (a) $k_- = 5p,\, k_+ = 5p$. (b) $k_- = 0.01p,\, k_+ = 0.002p$.}
 \end{figure}

Yet, the hole-based mean field theory is not sufficient to reproduce the CDR for slow defect turnover, Figs. \ref{J_rho_restr_fig}(c,d). Again, this is due to significant correlations between the defect site $i+1$ and the occupation of the left neighboring site $i$ (see Fig. \ref{corr_nn_rho_restr_fig}), which were neglected in the hole-based mean field approximation, Eqs. (\ref{hole-based_MF_eq1}) and (\ref{hole-based_MF_eq2}). Nonetheless, we can follow the same, particle-based, approach of the enhanced mean field theory as introduced for unconstrained dynamics (Eq. (\ref{mf_enhanced_eq1})) and consider the particle current in absence of exclusion interaction to obtain an approximate expression for $\langle \sigma_i(1-\nu_i) \rangle$. The waiting time is determined by the probability to encounter a defect. Now, however, we have to consider the real defect density of the constrained system, $\rho^{(c)}_d = k_+/(k_+ + k_-)(1-\rho)$. Thus, we can follow the same lines as for the enhanced mean field theory of the unconstrained model, by substituting $\rho_d^{(c)}$ in Eq. (\ref{tau_eq}) and obtain,
\bq
\label{mf_enhanced_constr_eq}
J \approx \frac{p}{1 + \frac{p}{k_-}\rho_d(\rho)}, \,\,\, \mbox{ with } \rho_d(\rho) = \frac{k_+}{k_+ + k_-}(1-\rho).
\eq
We see that this approximation correctly reproduces the skewness and shift of the maximum of $J(\rho)$ (Fig. \ref{J_rho_restr_fig}(c,d), red line) and gives a reasonable estimate for its magnitude.



\begin{figure}
\subfloat[]{\includegraphics[width=0.48\columnwidth]{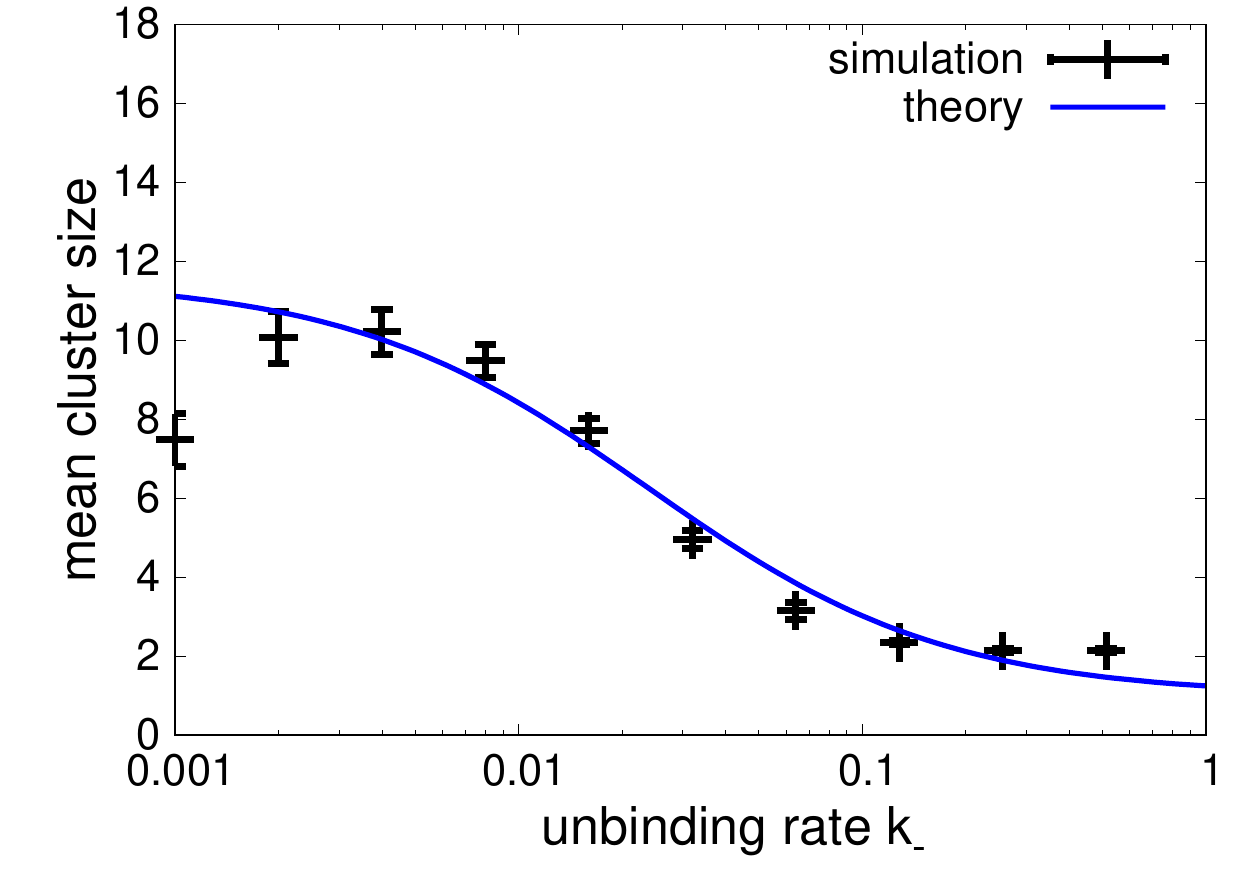}}
\subfloat[]{\includegraphics[width=0.48\columnwidth]{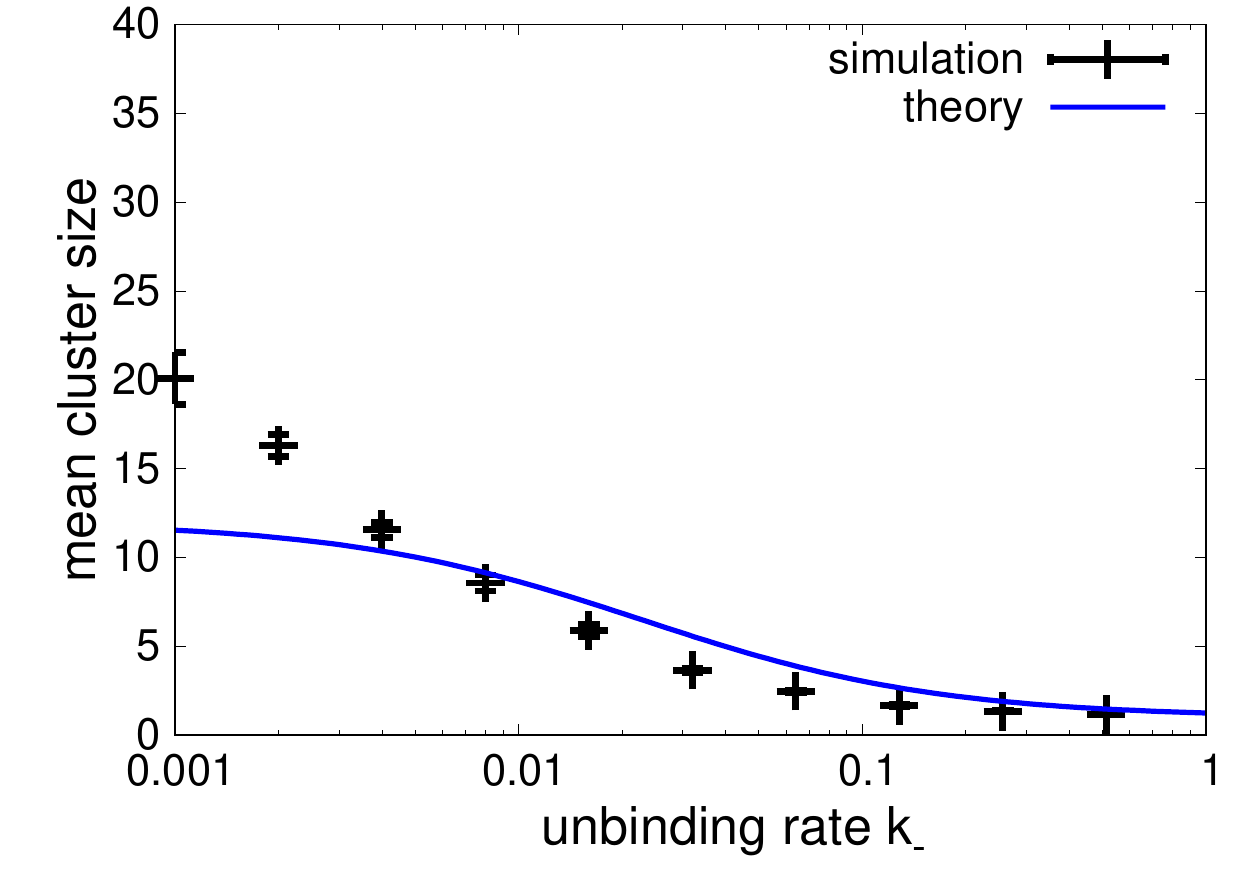}}
\caption{\label{av_cluster_restr_fig} Mean cluster size for constrained defect binding, 
for $T = 1000000/p$, $\rho=0.1$, $L=500$. (a) as a function of unbinding rate $k_-$, in units of $p$, for fixed $k_+ = 0.001p$. (b) as a function of $k_-$ while $k_+ = 0.1\, k_-$ is scaled so that the density of defects remains constant. The blue line is the theoretical estimate from Eq. (\ref{mean_clustersize_constr_eq}).
}
\end{figure}

Space-time plots in Fig. \ref{space-time_restr_plot}(c,d) show that the particle distribution becomes inhomogeneous for slow defects, which results in the formation of particle clusters. We can apply the theory of cluster initiation and growth (Section \ref{unconstrained_sec}) to estimate the mean cluster size. Similarly as for the unconstrained case, cluster growth is determined by two time scales, the life time $t_u = 1/k_-$ of an obstacle, and the time until a cluster ``saturates'' because trailing particles are cut off by a trailing defect, $t_e$. However, in the constrained case, the defect distribution is not necessarily equilibrated with respect to a cluster that has just been initiated, due to the particle-defect interaction. In order to find the cluster saturation time, we thus follow a different approach, which considers the cluster coagulation and de-coagulation dynamics and which is outlined in detail in Appendix B. There, we obtain the saturation time $t_e = \sqrt{\frac{\rho}{p\,k_+ (1-\rho)}}\rho_c^{-1}(1-\rho_c)^{-1}$. Hence, following the same line of arguments that led to Eq. (\ref{mean_cluster_eq}) for unconstrained dynamics, we find that the mean cluster size is
\bq
\label{mean_clustersize_constr_eq}
\bar l \approx J \bar t_{c} = \frac{p\rho_{c}(1-\rho_{c})}{k_- + \sqrt{\frac{p\,k_+ (1-\rho)}{\rho}}\rho_{c}(1-\rho_{c})}.
\eq
Figure \ref{av_cluster_restr_fig} compares the mean cluster sizes 
obtained in computer simulations and from Eq. (\ref{mean_clustersize_constr_eq}). The agreement is good, except for very low $k_-$ for which significant deviations are visible.

\subsection{Non-zero defect hopping rates, unconstrained defect binding}

\begin{figure}
\subfloat[]{\includegraphics[width=0.48\columnwidth]{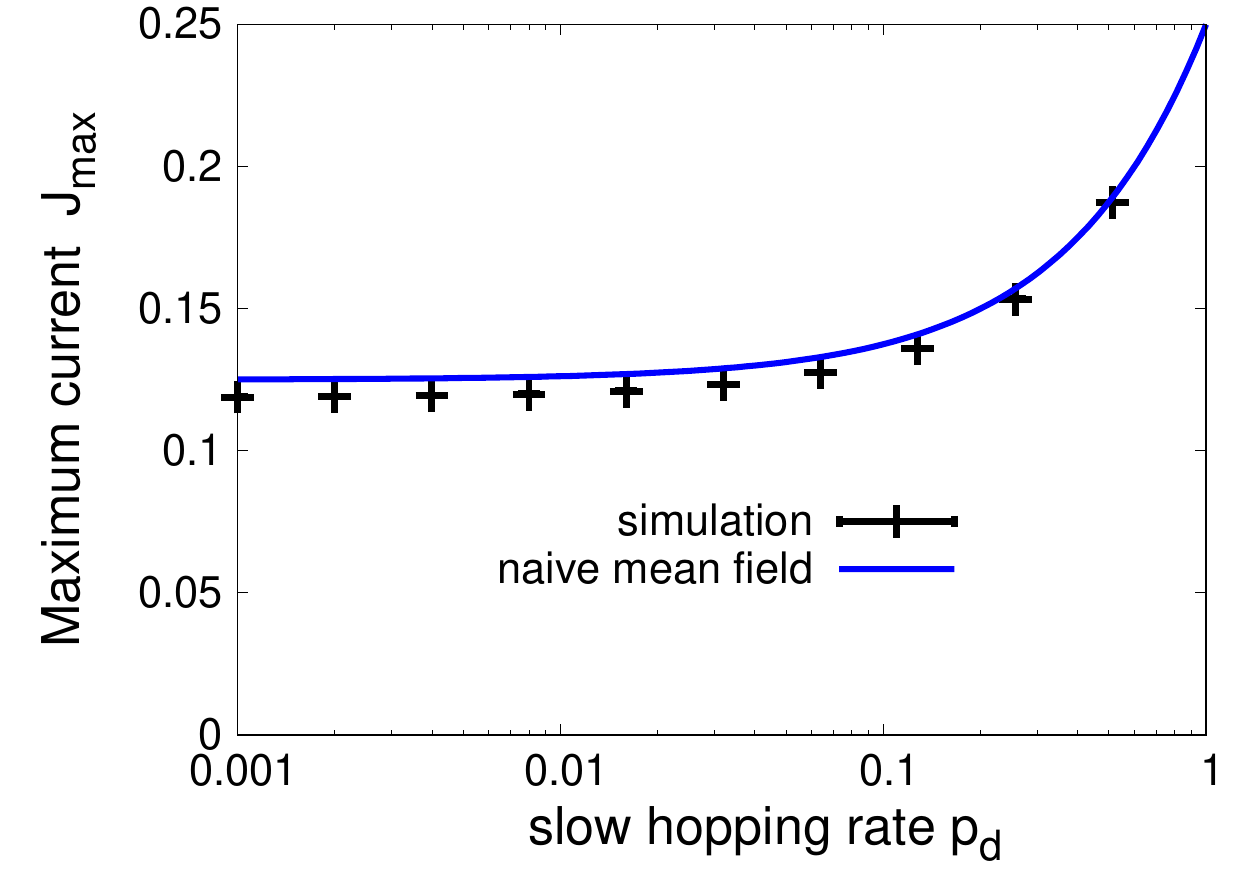}}
\subfloat[]{\includegraphics[width=0.48\columnwidth]{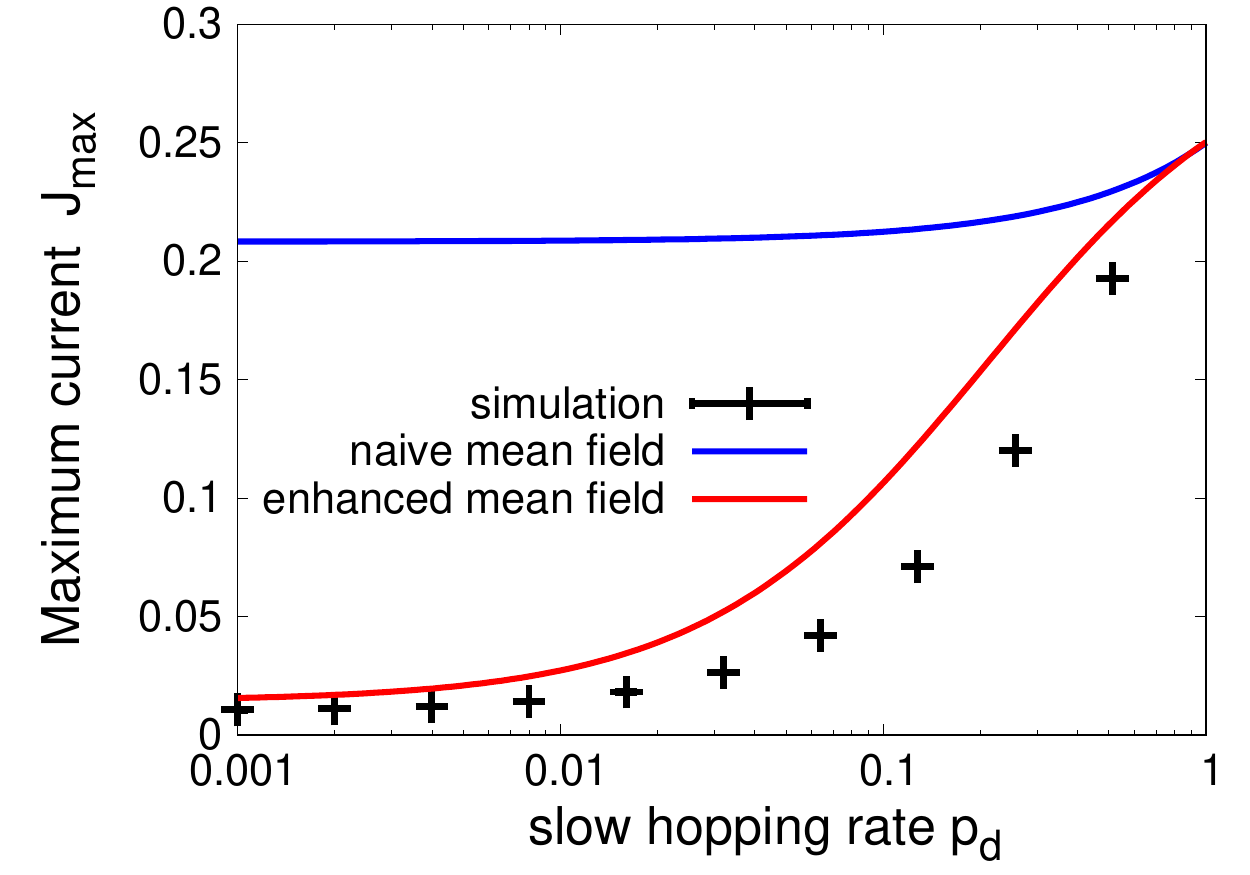}}
\caption{\label{J_pd_fig} Maximum current $J_{\rm max}$ as a function of slow hopping rate $p_{d}$ (both in units of $p$) for unconstrained dynamics and $L=1000, T=100000/p$ and (a) $k_- = 5p$, $ k_+ = 5p$ (b) $k_- = 0.01p, k_+ = 0.002p$. Data points are from simulations, the blue line is the naive mean field theory, Eq. (\ref{jmax_naive_pd_nonzero}) while the red line is the enhanced mean field theory according to Eq. (\ref{J_enh_MF_pd_eq}).}
\end{figure}

We can extend our theory for non-zero slow hopping rates $p_{d} > 0$. We shall only consider unconstrained dynamics for simplicity. Our ``naive'' mean field theory for fast defects trivially generalizes to  
\begin{align}
\label{naive_mf_pd_eq}
J \approx \left[\rho_{d} p_{d} + (1-\rho_{d}) \, p \right] \rho(1-\rho),
\end{align} 
and thus the maximum current at $\rho=1/2$ reads
\bq
	J_{\rm max} = (1/4)\left(\rho_{d} p_{d} + (1-\rho_{d}) \, p \right). \label{jmax_naive_pd_nonzero}
\eq
Figure \ref{J_pd_fig} shows the maximum current as a function of $p_d$ \footnote{Note that the CDR remains symmetric, which we do not show in a separate figure here.} . The theoretical prediction from Eq. (\ref{jmax_naive_pd_nonzero}) agrees very well with the simulation data for fast defect dynamics (panel a) but deviates significantly for slow defects (panel b).

To obtain an enhanced mean-field theory, in line with previous approaches in Sections \ref{constrained_sec} and \ref{unconstrained_sec}, we need to consider all possible ways in which a particle can hop to a new, particle-free site. The new site may be without obstacle (case A, probability $1-\rho_d$), or it may contain an obstacle (case B, probability $\rho_d$). In the first case A, the particle jumps with rate $p$. In the second case B, the particle either waits for the obstacle to unbind (rate $k_-$) and then jumps with rate $p\gg k_-$, or it jumps with rate $p_d$ with the obstacle still present at the arrival site. The total jump rate will be the sum of the rates of the latter two processes. In the limit $p \gg k_-,k_+,p_d$, the time to hop after defect unbinding, $1/p$ is negligible compared to the unbinding time $1/k_-$. Then the average waiting time for the jump to occur in absence of a particle on the next site (see section \ref{unconstrained_sec}) is
\bq
	\bar \tau_f \approx \rho_{d}\frac{1}{p_{d} + k_-} + (1 - \rho_{d}) \frac{1}{p},
\eq
where we have taken into account the probabilities of both scenarios A,B. This gives the following expression for the current
\bq
\label{J_enh_MF_pd_eq}
J = \bar \tau_f^{-1} \rho(1-\rho) \approx p \left[1 + \rho_{d}\left(\frac{p}{p_{d} + k_-} - 1\right)\right]^{-1} \rho(1-\rho) \,\,\, .
\eq
and the maximum current reads
\bq
\label{Jmax_enh_MF_pd_eq}
J_{\rm max} \approx \frac{p}{4} \left[1 + \rho_{d}\left(\frac{p}{p_{d} + k_-} - 1\right)\right]^{-1} \,\,\, .
\eq
We see in Fig. \ref{J_pd_fig}(b) that Eq. (\ref{Jmax_enh_MF_pd_eq}) gives a much better estimate for the maximum current for slow defect dynamics than Eq. (\ref{jmax_naive_pd_nonzero}).
Note that for $p_{d} = 0$ and for $p \gg k_-$ we recover the result (\ref{enhanced_mf_eq}).

\begin{figure}
\begin{center}
\includegraphics[width=0.8\columnwidth]{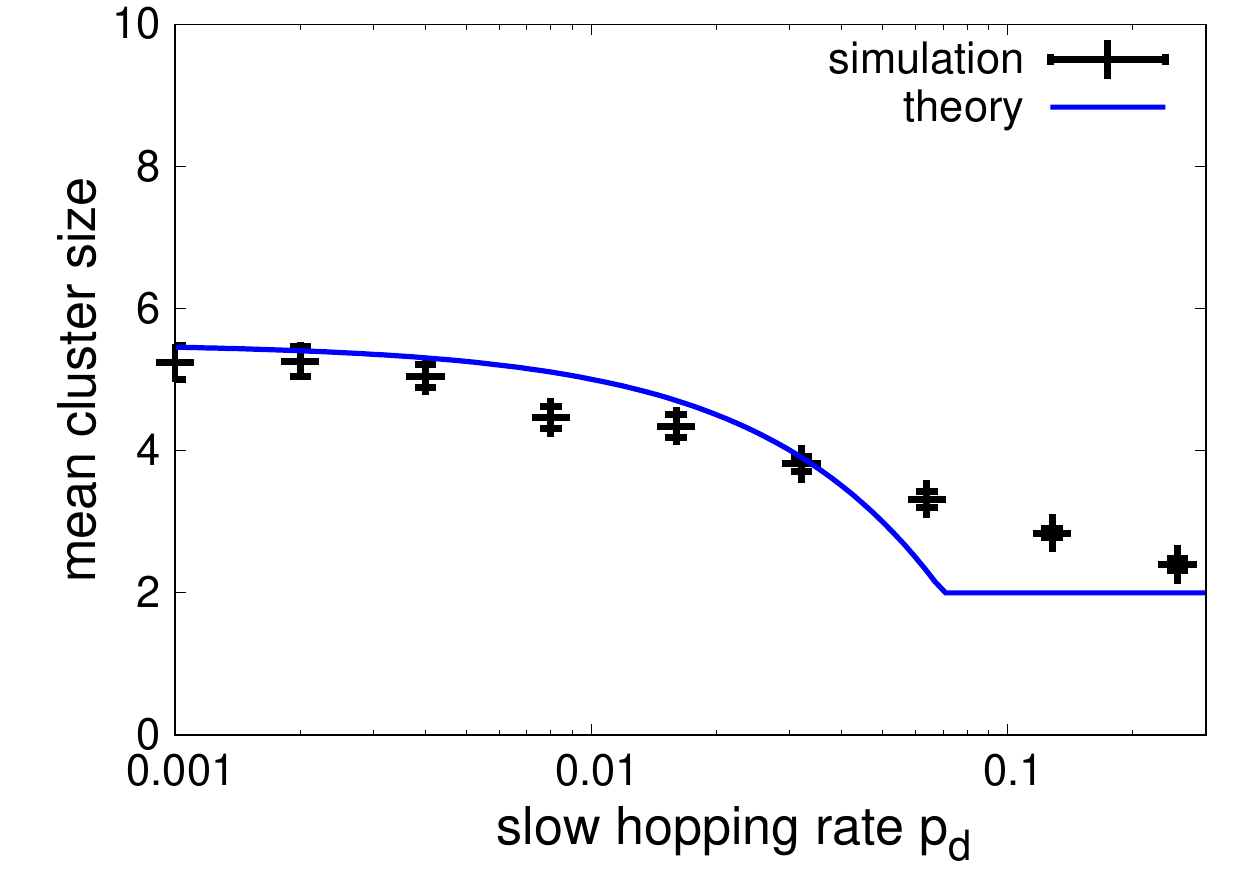} 
\end{center}
\caption{\label{av_cluster_pd_fig} Mean cluster size for unconstrained dynamics and $k_+ = 0.001p, k_- = 0.01p$, $T = 1000000/p$, $\rho=0.1$, $L=2000$, as a function of $p_{d}$, in units of $p$. The symbols are results from computer simulations, the blue line is the cluster growth estimate from Eq. (\ref{mean_cluster_pd_eq}).}
\end{figure}

To obtain the mean cluster size for $p_{d} > 0$ we need to take into account that a cluster not merely grows by incoming particles, but also shrinks as particles ``leak'' at the leading edge (defect site) with the slow hopping rate $p_{d}$. We thus need to subtract the leakage current $J_{\rm leak}$ from the incoming current of particles. We can assume that for $k_- \ll p$ the site immediately after a defect is unoccupied, so that the leakage current at a defect site $i$ of a cluster is $J_{\rm leak}\approx \rho_{i} p_{d}$. Apart from the loss of particles through the leading edge, particles can also be gained through the leakage at the trailing edge (nearest defect side behind the cluster). We have for the leading defect $\rho_{i} = 1$, whereas for the trailing defect at site $i-d$ the occupation probability is $\rho_{i-d} \approx \rho_{c}$. The total net current due to particles leaking in/out of the cluster is thus $\Delta J =\rho_{i}p_{d} - \rho_c p_{d} = p_{d}(1-\rho_{c})$. The particles lost through leakage during the time that the defect is bound, $J/k_-$, need to be subtracted from the cluster size. Note that the "leak" current can occur only if there is a cluster at all, while the expression derived here does not account for this condition. Considering that, by definition, a cluster must have at least two particles, we obtain following expression for the mean cluster size
\begin{align}
\label{mean_cluster_pd_eq}
\bar l &\approx \max\left[J \bar t_{c} - \Delta J/k_-,2\right] \\ \nonumber
&=  \max\left[\frac{p \rho_{c}(1-\rho_{c})}{p\rho_{c}\rho_{d} + k_-} - (1-\rho_{c})p_{d}/k_-,2\right].
\end{align}
Note that the time of leakage, $1/k_-$ may last longer than cluster growth in the absence of leakage. 

In Fig. \ref{av_cluster_pd_fig} we compare the theoretical curve from Eq. (\ref{mean_cluster_pd_eq}) with the results of computer simulations. Our theory matches the data reasonably well for very small $p_{d}$, while for larger $p_{d}$, when the cluster size approaches the trivial value of two (no clusters) the theory fails because our assumption $\rho_i = 1$ does not hold anymore.

\section{Open boundary conditions}
\label{open_bc_sec}

We now consider the unconstrained model\footnote{For the constrained model the arguments would follow the same lines but differ quantitatively.} with open boundary conditions (open BC). Particles enter the lattice at site $i=1$ with rate $\alpha$ and exit the lattice at site $L$ with rate $\beta$; no hopping from site L to site 1 can occur. The standard TASEP with open boundaries has three phases: A low-density phase, in which the density is determined by the entry rate $\alpha$, exists for $\alpha<p/2$ and $\alpha<\beta$. In a high density phase ($\alpha>\beta$, $\beta<p/2$) the current and particle density are determined by $\beta$. A third, maximum-current phase, in which the current becomes insensitive to the boundary rates exists for $\alpha>p/2$ and $\beta>p/2$.

The open boundary conditions can also be modelled by adding boundary reservoirs with fixed boundary densities $\rho_0$ and $\rho_{L+1}$ on virtual sites $i=0$ and $i=L+1$, attached to sites $i=1$ and $i=L$, respectively. The in- and outflow of particles from the lattice correspond to particles hopping from site $0$ to site $1$ and from site $L$ to site $L+1$ with the same hopping rate $p$ as the ``regular'' hopping rate. It has been shown that the phase diagram of a driven lattice-gas model (such as TASEP) can be obtained by looking at the extrema of the current density relation of the periodic system. This is known as the \emph{extremal current principle} \cite{kolomeisky_shockdyn,popkov2} and it states that
\begin{align}
J = \min_{\rho \in [\rho_{0},\rho_{L+1}]} J(\rho) \mbox{ for } \rho_{0} < \rho_{L+1}, \label{extr_curr_pr_eq} \\ \nonumber
J = \max_{\rho \in [\rho_{L+1},\rho_{0}]} J(\rho) \mbox{ for } \rho_{0} > \rho_{L+1} .
\end{align}
The virtual boundary densities $\rho_{0}$ and $\rho_{L+1}$ are in general directly related to the entry/exit rates, $\alpha$ and $\beta$, as shown below. In particular, for the TASEP, $\rho_{0} = \alpha/p$ and $\rho_{L+1} = (1 - \beta/p)$.

Equation (\ref{extr_curr_pr_eq}) shows that the structure of the phase diagram depends crucially on the number of maxima and minima in the current density relation of the corresponding model with periodic BC \cite{popkov2}. In particular, for $J(\rho)$ with a single maximum, as it is the case for our model, the phase diagram must have the same structure as the normal TASEP \cite{kolomeisky_shockdyn,popkov2}. Phase boundaries are determined by whether the current depends on the left boundary density $\rho_{0}$ (low density phase, LD), the right boundary density $\rho_{L+1}$ (high density phase, HD) or is independent of the boundary conditions (maximum current phase, MC),
\begin{align}
\begin{array}{llll}
J = J(\rho_{0})& \mbox{ for }& \rho_{0} < \rho_{\rm max}, \rho_{0} < 1 - \rho_{L+1}& \mbox{ (LD)}, \\ \nonumber
J = J(\rho_{L+1})& \mbox{ for }& \rho_{L+1} > \rho_{\rm max}, \rho_{0} > 1 - \rho_{L+1}& \mbox{ (HD)}, \\ \nonumber
J = J_{\rm max}& \mbox{ for }& \rho_{1} > \rho_{\rm max}, \rho_{L+1} < \rho_{\rm max}& \mbox{ (MC)},
\end{array}
\end{align}
where  $\rho_{\rm max}$ is the density for which $J$ is maximised ($J(\rho_{\rm max})=J_{\rm max}$), which is $\rho_{\rm max} = 0.5$ for the unconstrained ddTASEP.   Here, we utilised the symmetry of the CDR, $J(\rho) = J(1-\rho)$. 

\begin{figure}[p]
\subfloat[]{\includegraphics[width=0.48\columnwidth]{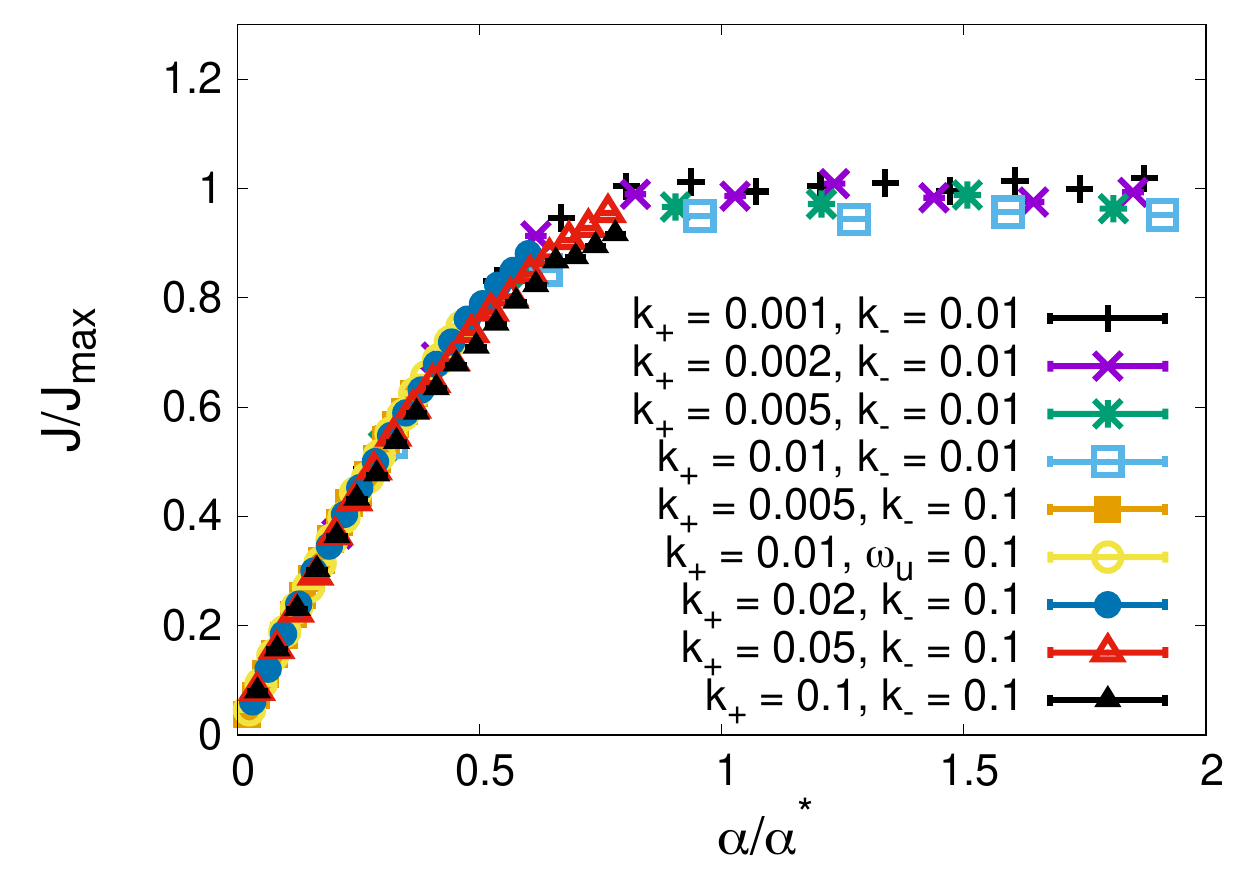}}
\subfloat[]{\includegraphics[width=0.48\columnwidth]{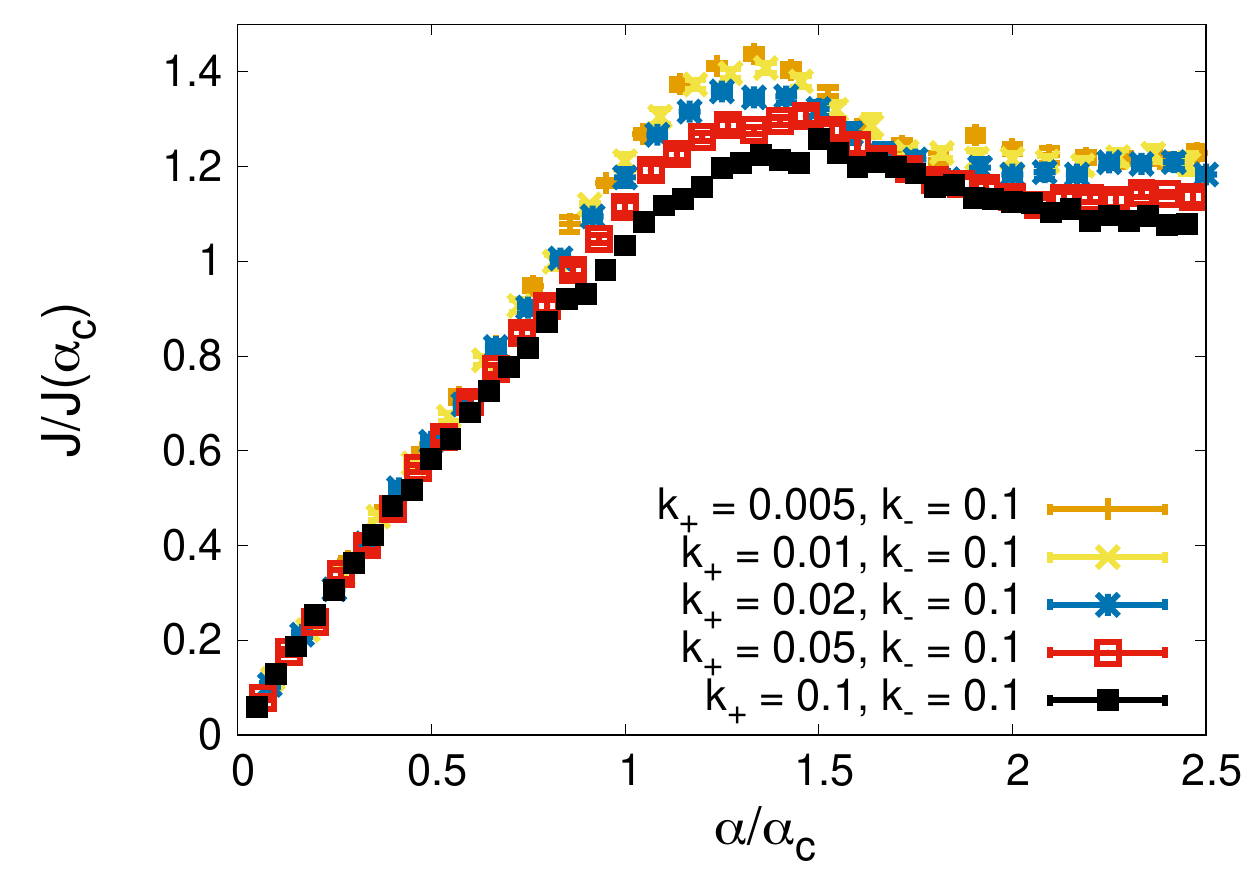}} \\
\begin{center}
\subfloat[]{\includegraphics[width=0.61\columnwidth]{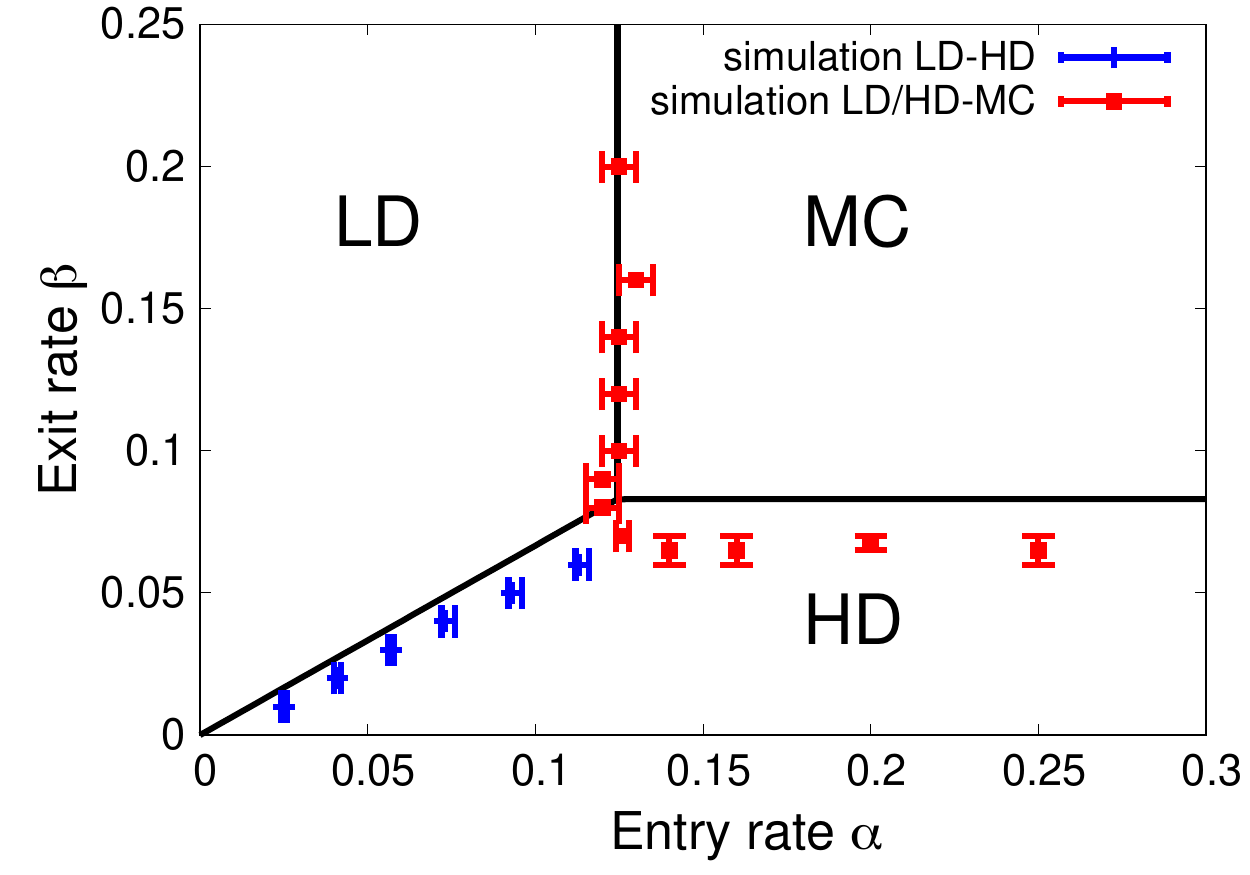}}
\end{center}
\caption{\label{J_alpha_fig} Phase transitions in the open, unconstrained ddTASEP. (a,b) Renormalised current for $L=1000, T=100000/p$ and various values of $k_-$ and $k_+$ (see legend, in units of $p$). (a) Renormalised current $J/J_{\rm max}$ ($J_{\rm max} = J(\rho=1/2)$ in the periodic system) as function of $\alpha/\alpha^*$ for $\beta = p$, (b) Renormalised current $J/J(\alpha=\alpha_c)$ as function of $\alpha/\alpha_c$ for $\beta = 0.02p$. (c) Phase diagram for $k_- = 0.1p, k_+ = 0.05p$ ($\rho_d = 1/3$), with the low density phase (LD), high density phase (HD), and maximum current regimes (MC) marked correspondingly. On the axes are entry and exit rate, respectively, in units of $p$. Bold lines are the results from the mean field theory, Eqs. (\ref{alpha_c_eq}), points are numerically (computer simulations) determined phase boundaries. Blue crosses mark the transitions between HD- and LD-phase, while red squares are second order transitions towards MC-phase. The LD-HD transition has been identified by increasing $\alpha$ from zero to $0.3p$ in small steps and recording the value of $\alpha$ at which the particle density crossed $\rho=0.5$ for the first time. The LD-MC transition has been found by increasing $\alpha$ and recording the value at which the current reached $J_{\rm max} = J(\rho=0.5)$ (determined in simulations of the periodic system) for the first time. The HD-MC transition has been found in a similar way by increasing $\beta$ from zero to $0.25p$. Error bars denote upper and lower bounds, marked by the first crossing of $\rho=0.5$ by $\rho + \Delta \rho$ respectively $\rho - \Delta \rho$ or by first reaching $J_{\rm max}$ for $J + \Delta J$, where $\Delta \rho, \Delta J$ are standard errors of mean from simulations (10 replica).}
\end{figure}

In the following we take the continuous limit (valid for large system size $L$) for which we can approximate the boundary densities as $\rho_0 \approx \rho_1$ and $\rho_{L+1} \approx \rho_L$. These densities can then be determined from the continuity equation for boundary currents via a mean field approximation. In the steady state ($\partial_{t} \rho_{i} = 0$), the continuity equations for sites $i=1$ and $i=L$ read:
\begin{align}
0 &= \alpha \langle (1-\nu_{1}) (1- \sigma_{1}) \rangle - \langle p(1-\nu_{2}) \sigma_{1} (1-\sigma_{2}) \rangle \\ \nonumber
&\approx \alpha (1-\rho_{d})(1-\rho_{1}) - 4 J_{\rm max} \rho_{1}(1-\rho_{1}), \\ \nonumber
0 &= p \langle (1-\nu_{L}) \sigma_{L-1} (1- \sigma_{L}) \rangle - \beta \langle \sigma_{L} \rangle \\ \nonumber
&\approx 4 J_{\rm max}\rho_{L}(1-\rho_{L}) - \beta \rho_{L},
\end{align}
where we employed a mean field approximation and approximated the current as $J \approx 4 J_{\rm max} \rho(1-\rho)$ (note that $J_{\rm max} = J(\rho = 1/2)$). From this it follows that
\begin{align}
\label{bound_res_eq}
\rho_0\ \approx \rho_{1} &\approx \frac{\alpha (1- \rho_d)}{4 J_{\rm max}} , \\ 
\rho_{L+1} \approx \rho_{L} &\approx 1 - \frac{\beta}{4 J_{\rm max}}.
\end{align}
Thus, by defining the critical entry rates $\alpha_{c} := \beta/(1- \rho_{d})$ and $\alpha^{*} := 2 J_{\rm max}/(1 - \rho_d)$, as well as the critical exit rate $\beta^{*} := 2 J_{\rm max}$,  the phase boundaries, according to the extremal current principle \cite{popkov2} are
\begin{align}
\label{alpha_c_eq}
\alpha &< \alpha_{c}, \alpha < \alpha^{*} \mbox{ (LD)} \\ \nonumber
\alpha &> \alpha_{c}, \beta < \beta^{*} \mbox{ (HD)} \\ \nonumber
\alpha &> \alpha^{*} \beta > \beta^{*} \mbox{ (MC)}
\end{align}
In particular, in the low- and maximum-current phase ($\alpha<\alpha_c$) the current is given by
\bq
	J = \left\{ \begin{array}{ll}
		\alpha(1-\rho_d)\left(1-\frac{\alpha(1-\rho_d)}{4J_{\rm max}}\right) & \mbox{for}\; \alpha<\alpha^* \\
		J_{\rm max} & \mbox{for}\; \alpha\ge \alpha^*
	\end{array}
	\right.  .\label{eq:current_LD}
\eq
We note that, in contrast to the TASEP with quenched site-wise disorder \cite{asep_dis} the maximum current does not depend on the system size $L$. Furthermore, when we express the current as a function of the re-scaled parameter $\hat \alpha := \alpha/\alpha^*$, we get a universal, parameter-free expression for the current in the LD phase,
\bq
	\frac{J}{J_{\rm max}} = \left\{ \begin{array}{ll}
		\hat \alpha (1 - \hat \alpha) & \mbox{for}\; \hat \alpha<1 \\
		1 & \mbox{for}\; \hat \alpha\ge 1
	\end{array},
	\right.  \label{eq:J_alpha_scalefree}
\eq
provided that $\beta$ is large enough to evade the transition to the high-density phase. Equation (\ref{eq:J_alpha_scalefree}) predicts that if we take the current $J$ obtained from simulations for different parameters $\beta,k_-,k_+$, divide it by $J_{\rm max}$, and plot it as a function of $\hat \alpha = \alpha/\alpha^*$, all curves $J(\alpha)$ should `collapse' onto a single, universal curve. To test this prediction we plot the current as function of $\alpha/\alpha^{*}$ and $\beta = p$ in Fig. \ref{J_alpha_fig}(a) for different values of $k_{+,-}$. According to the extremal current theory, Eq. (\ref{alpha_c_eq}), a transition from a regime in which $J$ depends on $\alpha$ towards a regime where $J$ is independent of $\alpha$ should occur at $\hat \alpha=1$. This is indeed what Fig. \ref{J_alpha_fig}(a) shows. We thus conclude that the extremal current principle is able to identify the phase transitions correctly in our model. 

Another scaling relationship can be found using the scaling parameter $\tilde \alpha = \alpha/\alpha_c$, yielding $J/J_c = \tilde \alpha (1 - \tilde \alpha \beta/4 J_{\rm max})$ with $J_c = J(\tilde \alpha = 1)$. In this form, however, the rescaled current still depends explicitly on $\beta$ and $J_{\rm max}$. This scaling relationship applies only when the exit rate is small, and is shown in Fig. \ref{J_alpha_fig}(b) for $\beta=0.02$. Notably, the curve $J/J_{\rm max}$ displays a ``bump'' where it reaches the maximum, i.e. where the transition between LD- and HD-phase is expected. The mean field approach presented here cannot explain this bump. We hypothesize that the bump may be caused by the de-confinement of shocks at the transition point \cite{ludger_asep_shockfluct}, which introduces strong correlations. Modelling such correlations will, nonetheless, require going beyond the mean-field framework presented in this work. 

Figure \ref{J_alpha_fig}(c) shows the full phase diagram of our model. Phase boundaries obtained from the mean-field theory (represented by lines) estimate well the results of computer simulation (see the figure caption for how the phase boundaries have been determined numerically). Deviations are due to the approximative nature of the mean field approach in Eq. (\ref{bound_res_eq}) used to obtain $\alpha_c,\alpha^*$.

\section{Application to gene transcription}
\label{application_sec}

We shall show how our results can be used to explain a curious biological observation. Transcription is a process in which biological cells make mRNA from a DNA template. DNA is a long polymeric molecule made from four monomers (adenine, thymine, guanine, cytosine, abbreviated as A,T,G,C) called nucleotides. 
In order to produce proteins, DNA must be first transcribed onto another linear polymeric molecule, the mRNA 
\cite{alberts_molecular_2002}. Transcription is effectuated by a molecular machine -- the RNA polymerase -- which attaches to a special DNA sequence (transcription start site) and begins to proceed along the DNA, ``reading off'' the DNA sequence and adding appropriate nucleotides to a newly created mRNA chain. The polymerase detaches from the DNA when it encounters another special sequence (transcription end site). 

We can immediately see analogies between transcription and TASEP as they both involve particles moving along a one-dimensional chain. In fact, modelling transcription was the motivation behind the very first TASEP paper \cite{macdonald_kinetics_1968}. More specifically, transcription initiation, elongation, and polymerase detachment correspond to TASEP particles entering site 1, moving along the chain, and exiting at site $L$, respectively.

The original application of TASEP to transcription did not involve any obstacles. However, we now know that DNA forms a highly dynamic, three-dimensional structure, with many proteins transiently bound to it. Such proteins can be transcription factors whose binding sites occur in many different places on the DNA \cite{lin_transcriptional_2012,lovelace_regulatory_2016}, or histones around which the DNA is wrapped and which are known to impede transcription \cite{li_role_2007,workman_alteration_1998}.

We shall now show that our model with dynamic obstacles can explain recent experimental results. It has been shown in Ref. \cite{veloso_rate_2014} that the speed with which RNA polymerases move along the DNA and the rate with which mRNA is produced depend on certain genomic features. We are particularly interested in two such features: DNA methylation (fraction of cytosines that have an additional methyl group attached) and CG density (the number of cytosine-guanine dinucleotides per 1000 nucleotides of single-stranged DNA). DNA methylation is known for its regulatory effects on transcription \cite{jones_functions_2012}, whereas CG density probably does not directly affect transcription but it correlates with methylation density. In what follows we shall use CG density as a proxy for DNA methylation density since the latter quantity is much more difficult to measure.

\begin{figure*}
\begin{center}
	\includegraphics[width=0.7\columnwidth]{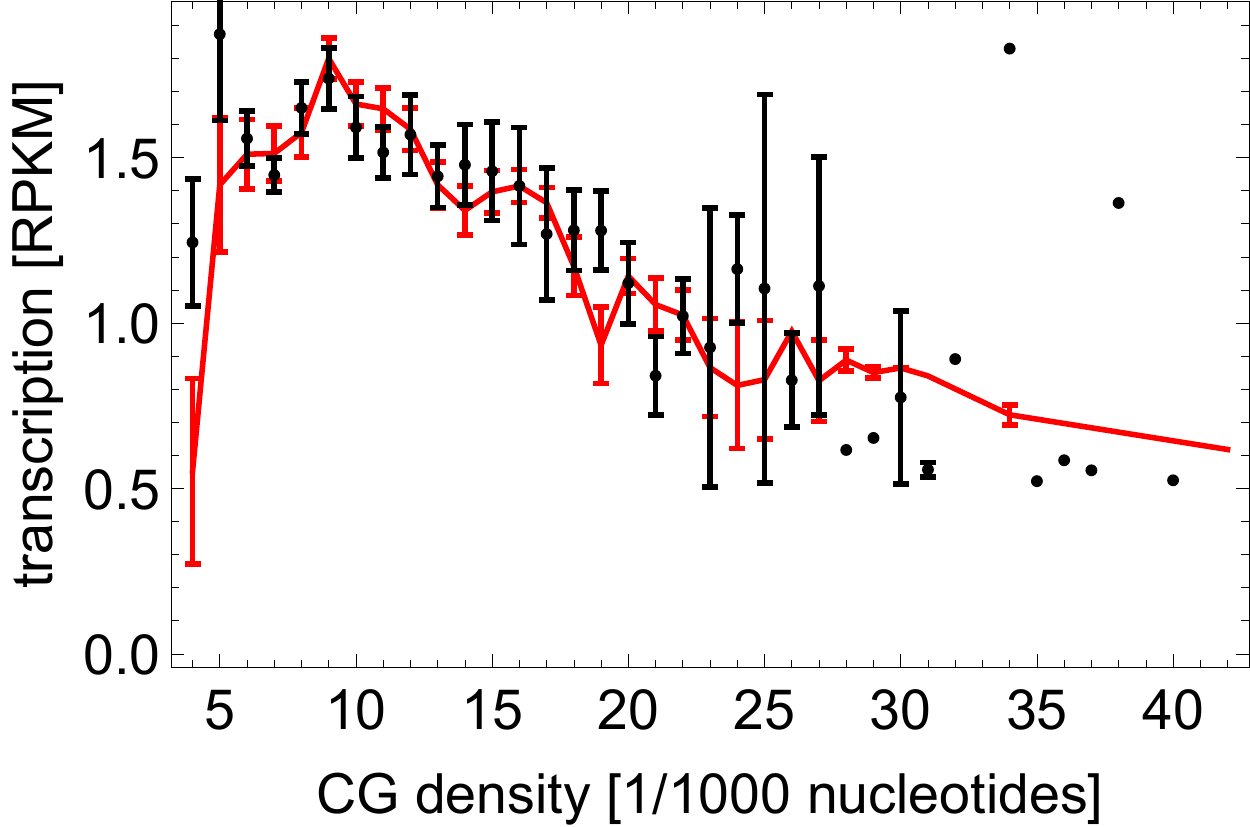}
\end{center}
\caption{\label{J_transcription}Transcription rate (RPKM, see Appendix A) versus CG density. Black = experimental data for K562 human leukemia cells \cite{veloso_rate_2014}. Red = model predictions for $C/k_-=1/0.09$ and gene-dependent initiation (entry) rates $\alpha$ from Ref. \cite{veloso_rate_2014}.}
\end{figure*}

Figure \ref{J_transcription}  (black points) shows the experimentally measured transcription rate versus CG density  for a particular cell line from Ref. \cite{veloso_rate_2014}, see Appendix A for details. To make this plot we binned genes according to their CG density (bin width = 1/1000 nucleotides) and calculated the mean and its standard error in each bin. 
Clearly, transcription slows down with increasing CG density.

To explain this, we hypothesize that the RNA polymerase is slowed down by obstacles that bind to the DNA. We assume that one site in our model corresponds to 60 nucleotides of the DNA because this is the size of RNA polymerase (one polymerase = one particle in the model). The maximum speed of the polymerase calculated from Ref. \cite{veloso_rate_2014} is 55 nucleotides/s. We therefore take $v=1/s$ ($55/60\approx 1$) as the obstacle-free hopping rate, and use formula (\ref{eq:current_LD}) to predict the rate of transcription for each gene. We assume that $J_{\rm max}$ in (\ref{eq:current_LD}) is given by Eq. (\ref{enhanced_mf_eq}) for $\rho=1/2$, that is
\bq
	J_{\rm max} = \frac{v}{4(1+Cv\rho_{CG}/k_-)}.
    \label{pred_trans_eq}
\eq 
The proportionality parameter $C$ is used to convert CG density $\rho_{\rm CG}$ to density of defects $\rho_d$. In particular, $C/k_-$ can be interpreted as the fraction of CG sites occupied by obstacles, divided by the unbinding rate of the obstacle. This parameter, as well as the unknown proportionality factor $F$ in gene expression $= F \times J$ are the only unknown parameters that must be fitted to data (see also Appendix A). The remaining input parameters are the initiation rates ${\alpha}$ and CG densities ${\rho_{CG}}$ of individual genes which we take from Refs. \cite{veloso_rate_2014} and \cite{fuchs_2014}. 

Figure \ref{J_transcription} (red line) shows the best-fit transcription rate (averaged over many genes as described above) to experimental data. Note that since genes in different bins may have different initiation rates (known from Ref. \cite{veloso_rate_2014}), the theoretical curve appears `wiggly'. The best-fit value of $C/k_-$ is $1/0.09$s. While it is not possible to determine the values of $C$ and $k_-$ from the ratio $C/k_-$ alone, we can estimate $k_-$ if we assume a certain density of obstacles. For example, if we take that 50\% of CGs are occupied by obstacles ($C=0.5$), $k_-=0.09 \times 0.5=0.045$s$^{-1}$ and the mean life time of obstacles reads $\tau_d=1/k_-\approx 22$s, which is typical for many DNA-binding proteins \cite{klose_dna_2005,lever_rapid_2000}. While we cannot unambiguously identify the nature of the obstacles, our calculation shows that transcription slow-down due to dynamic disorder seems to be a plausible biological mechanism.

\section{Conclusions}

In this work we study a version of the totally asymmetric exclusion process with dynamic disorder (ddTASEP) in which defects, which slow down the movement of particles or block it completely, appear and disappear randomly on any site. This is motivated by the binding and unbinding of proteins in intracellular transport and DNA transcription, which serve as obstacles to transport, but may also apply to various traffic scenarios in which dynamic obstacles are present (e.g. traffic lights).  

We consider two versions of this model, (i) when obstacles appear and disappear independently of particle occupation, (ii) when obstacles can only appear on empty sites. 

For periodic boundary conditions we investigate properties of the current-density relation (CDR), i.e., the current as function of the particle density. We perform computer simulations of the model and observe that for unconstrained defect dynamics the symmetric, parabolic form of the CDR of the standard TASEP is preserved, while for constrained defect dynamics, the CDR becomes skewed for slow defect turnover ($k_+,k_- \ll p$). We also observe a spatially heterogeneous distribution of particles for slow defect dynamics, in particular the formation of large particle clusters. 

To understand the results of computer simulations we develop a range of mean-field approaches of increasing complexity. These enable us to derive analytic estimates for the CDR and mean particle cluster size. These approximations reproduce well the magnitude and features of the CDR, for example the skewness/symmetry for constrained/unconstrained defect dynamics, and of the mean cluster size for varying defect (un-)binding rates.

We also study the model with open boundaries in which particles enter on one end of the lattice and exit on the other. We use an extremal current principle to show that the model exhibits the same phases as the standard TASEP but with altered phase boundaries, which are well-approximated by our mean field theory. 

Dynamic defects, in the form of proteins binding to DNA or structural features of chromatin, have been recently recognized as an important determinant of gene transcription. We show that the ddTASEP is able to explain why gene transcription depends on certain genomic features such as CG density and methylation. Our hypothesis is that proteins that bind to these DNA features or chromatin modifications act as obstacles for transcription and block the RNA polymerase -- a molecular machine which moves along DNA and produces mRNA.

Besides transcription, we expect that other intra-cellular processes such as transport by motor proteins can be affected by dynamic defects. For example, microtubule-associated proteins which bind to microtubules  may obstruct the progress of kinesin and dynein motors. 
A crucial difference to our model is that motor proteins themselves can (un-)bind from/to transport filaments. This has been modelled by TASEP variants which do not conserve the number of particles, such as the TASEP with Langmuir kinetics \cite{pff1}. Static defects \cite{pff_dis,pff_1def} and dynamically disordered binding rates \cite{ludger_pff-binding-disorder} have been considered in previous works, yet disorder has not been discussed in terms of obstacles (slow sites). It would be illuminating to see whether the main conclusions for our model (current-density relation, TASEP-like phase diagram, emergence of clusters in the low-density phase) remain true for a ddTASEP with Langmuir kinetics. 

Our work closes a substantial gap in the field of driven diffusive systems. While the TASEP with quenched disorder \cite{barma1,barma2,barma_driven_2006,asep_def,asep_dis}, isolated dynamic defects \cite{klumpp_dyn-def_2014,klumpp_dyn-def_2016,turci_transport_2013,das_particles_2000}, and disorder with particle-induced unbinding \cite{bus-route} has been studied before, we study for the first time the TASEP with random dynamic disorder. Although we do not progress beyond the mean field theory, we can reproduce many features (magnitude, skewness) of the CDR obtained from computer simulations. It remains an open question whether our model can be solved exactly as in the case of the ordinary TASEP. We think that the persistence of the parabolic shape of the CDR (very much like the CDR of the ordinary TASEP) in the unconstrained version of model may hint towards some hidden symmetries of its steady-state configurations. Finding such symmetries, and exploring connections between this model and zero-range-like processes (cf. Appendix B) will be an interesting future research project.

\section*{Acknowledgments}
We thank Luca Ciandrini for help with literature research. B.W. was supported by an RSE Personal Research Fellowship.





\section*{Appendix A}
We calculated CG density by counting CG pairs for each gene from the human hg19 reference genome data (GRCh37.74), and dividing by the length of gene. We took gene expression levels from Supplementary Table 1, Ref. \cite{veloso_rate_2014} (units: RPKM, Reads Per Kilobase of transcript per Million mapped reads). RPKM measure the amount of mRNA from a given gene accumulated in the cell, not the actual transcription rate $J$. However, if we assume that mRNA is degraded with (possibly gene-dependent) constant rate $d$, we expect gene expression level to be proportional to $J/d$. Assuming further that degradation rates for different genes are uncorrelated, and averaging over many genes (see below) we obtain that $J\propto$ RPKM.

To plot gene expression versus CG density, we took pairs (CG density, expression in RPKM) for all genes for which expression had been measured, and binned them according to CG density. Bin 1 contained all genes with CG density between 0 and 1/1000 nucleotides, bin 2 contained genes with CG density between 1/1000 and 2/1000 nucleotides, etc. For each bin we calculated mean expression and its standard error.

To predict gene expression using our model (Eqs. (\ref{eq:current_LD}) and (\ref{pred_trans_eq})), we took initiation rates from Additional File 6, Ref. \cite{fuchs_2014} (units: 1/min). We then calculated theoretical expressions for each gene from the data set for which we knew its initiation rate as
\begin{align}
	{\rm RPKM}_{\rm theor} &= N \left\{ \begin{array}{ll}
		\alpha(1-\alpha/(4J_{\rm max})), & \alpha<2J_{\rm max} \\
		J_{\rm max}, & \alpha\ge 2J_{\rm max}
	\end{array}
	\right. \\
    J_{\rm max} &= \frac{v}{4(1+Bv\rho_{\rm CG})},
\end{align}
where $N,B$ were two unknown parameters ($B=C/k_-$). We note that here we replaced $\alpha(1-\rho_d) \to \alpha$ compared to Eq. (\ref{eq:current_LD}), since the measured initiation rate corresponds to RNA polymerases that actually 'enter' the DNA, i.e. in absence of defects at the initiation site. We binned the genes as for the experimental data, and found best-fit $N,B$ that minimized the sum of squared differences between the binned ${\rm RPKM}_{\rm theor}$ and experimental RPKMs for CG densities between 0 and 25/1000 nucleotides:
\bq
	S=\sum_{\rho=0}^{25} \frac{({\rm RPKM}_{\rm theor}(\rho) - {\rm RPKM}_{\rm exp}(\rho))^2}{{\rm SE}_{\rm theor}(\rho) {\rm SE}_{\rm exp}(\rho)},
\eq
where SE denotes standard error of RPKM.

\section*{Appendix B}

Here we determine the saturation time $t_e$ for clusters growing under constrained defect dynamics, Eq. (\ref{model_varB}). The saturation time is the time needed for all particles between two defects to accumulate in a continuous queue, in absence of defect unbinding, $k_- \to 0$. The queue length in this limit corresponds to the total number of particles between two defects (as for unconstrained dynamics, see Section \ref{unconstrained_sec}). To approach this problem, we first consider the mapping to a version of the totally asymmetric zero-range process (ZRP) \cite{review_ZRP} with site-wise dynamic disorder, which we call ddZRP. The totally asymmetric ZRP is a lattice model in which each site can carry an arbitrary amount of particles, which can hop from site $i$ to site $i+1$ with a rate that only depends on the number of particles on the current site $i$, but not on that of any other site \cite{review_ZRP}. Furthermore, for the disordered case, it may depend on the defect state of each ZRP site $i$, $\nu_{i} = 1,0$. We map the constraint ddTASEP on the ddZRP by identifying each hole in the ddTASEP (ordered from left to right) as site $\tilde i$ in the ddZRP (i.e. $\tilde i$ corresponds to the $i$-th hole counted from the left) and the particles left of this hole, as the particles on site $\tilde i$, so that the particle number on each ddZRP site, $n^{ZRP}_{\tilde i}$, corresponds to consecutive stretches of particles, i.e. clusters. Since ddZRP sites correspond to ddTASEP holes, the defect dynamics on ddZRP sites are unconstrained. We note that the ddZRP has a different system size, $L_{ZRP} := (1-\rho)L$ (the number of holes, which is conserved). 


We can now consider cluster dynamics as a \emph{coagulation-decoagulation (CD) model}, as studied in Ref. \cite{benAvraham_coag_JStatMech1990}. In this view, for simplicity we consider all particles between two defect sites (in the ddTASEP) as a cluster (which is true for most of the time for $k_- \ll p$, when all particles accumulate in a queue). 

A cluster moves forward whenever a defect unbinds, with rate $k_-$. Since after each unbinding event a cluster moves on average $\bar d_{ZRP} := 1/\rho_d$ ZRP sites, this corresponds to a random walk in the variable $\tilde i - tk_-$ ($t$ = elapsed time) with diffusion constant $D = (\bar d_{ZRP})^2 k_-/2$. Two clusters coagulate, forming a single cluster, if any defects between them disappear, i.e. when a cluster moves onto another cluster. As long as all particles in a cluster (all particles between two defects) accumulate on a single site (in the ddZRP), no defects can bind between those particles to separate the cluster, thus it cannot \emph{de-coagulate}. Only when the cluster moves forward, at rate $k_-$, the particle of a cluster stretches out over several sites, and defects may bind between particles of a cluster, leading to the de-coagulation of the cluster. The distance to the next defect is on average $\bar d_{ZRP} = 1/\rho_c$ sites. During this process, particles move with speed $p$ between two defects, thus it takes an average time of $\bar t_d = \bar l/p$, where  $\bar l$ is the cluster size, until the particles have piled up again on a single site. During that time period defects can bind on the sites between the previous and the next defect, thereby separating the cluster. The cluster decoagulation probability is therefore approximated by the defect binding rate $k_+$, times the number of sites between the initial defect and the next one, $1/\rho_d$, times the time it takes for all cluster particles to reach it, $\bar l/p$, which is $k_+ \bar l/p\rho_d$. Thus the de-coagulation rate is $\omega_D = \frac{k_- k_+ \bar l}{\rho_d p}$. According to Ref. \cite{benAvraham_coag_JStatMech1990}, the equilibrium density of clusters of such a coagulatio/de-coagulation process is $\rho_{CD} = \omega_D\bar d_{ZRP}/2D$, and thus,
\begin{align}
\rho_{CD} = \frac{\omega_D \rho_d}{k_-} = \frac{\bar l k_+}{p}.
\end{align}
The number of clusters, $n_c = \rho_{CD} L_{ZRP}$, is related to the mean cluster size $\bar l$ by $n_c \bar l = N = \rho L$. Summarising this, we have
\begin{align}
\frac{\rho}{(1-\rho)\bar l} &= \frac{n_c}{L(1-\rho)} = \rho_{CD} = \frac{\bar l k_+}{p} \\
\Leftrightarrow \bar l &= \sqrt{\frac{p \rho}{k_+ (1-\rho)}}.
\end{align}
With $J t_e = \bar l$, and $J = p\rho_c(1-\rho_c)$ as for unconstrained dynamics (see Section \ref{unconstrained_sec}), we thus obtain 
\begin{align}
t_e = \frac{\bar l}{J} = \sqrt{\frac{\rho}{p \, k_+ (1-\rho)}}(\rho_C(1-\rho_C))^{-1},
\end{align}
which is used in section \ref{constrained_sec}.

\bibliography{references}  


\end{document}